**Dynamic Distributed Mobility Management System based on Multiple Mix-Zones over Road Networks**

***Imran Memon*[*1] , *Qasim Ali Arain*[2]**

[*1]College of Computer Science ,Zhejiang university ,Hangzhou,310027,Zhejiang ,China

[2]Beijing University of Posts and Telecommunications,Beijing 100876, People's Republic of China
E-mail: imranmemon52@zju.edu.cn

**Abstract**

Vehicle have access to the internet for communications to facilitate the need of mobility management and point of interest distribution in emerging Intelligent Transportation System (ITS) . Therefore, its obvious that by changing the road side unit frequently , may require fast handover management to mobile user (MU) for the configuration of new IP address , however the ongoing session remain un-disturbed . Recent study has shown that , the current version of IP mobility protocols are the centralized solutions, In centralized environment data traffic and data management is being routed to an anchor entity.In some situations , the vehicles may be routed in the form of group to pass from one RSU to another RSU.In response the traffic and mobility exchanged messages are increased and it will dramatically affect the performance of network.To cope with these challenges a new construct named DDMM (Dynamic Distributed Mobility management) has been anticipated by IETF DDMM organization. It has been designed on the basis of network-based CMM protocol, which is known as Proxy Mobile IPv6(PMIPv6).However , it has been realized that there can be significance difference among network based DDMM and PMIPv6 with respect to handover latency and packet loss.Therefore , we have envisioned fast handover for network-based DDMM (Dynamic Distributed Mobility management) which is based on fast handover for PMIPv6(PFMIPv6).However , it has been required to include modifications on PFMIPV6 to take on DDMM. We have designed some important additions to support this model, when MU hold IP flows and has many different anchor entities.In addition we also derived analytical expressions to check and made comparison about the handover performance of the given DDMM schemes.It has been revealed that DDMM has outperformed in terms of handover latency , session recovery and packet loss as compared with previous schemes.

## 1. Introduction

The recent progress in the field of mobile user devices has shown remarkable results and led to violent usage of network resources and has enlarged the traffic load. Apart from that, it has been studied that wireless technologies for the next generation networks will be the unique networks which might based upon IP technology. It is assumed that it might be 55% by 2017. It has been studied that , during vehicular communication multiple factors are responsible for the network peroformance , that might include; building ,trees , multiple intersections and road traffic information[1].Moreover it is assumed that high speed and the plenty of mobile users (MUs) in the road may be responsible for affecting the network performance because of frequent handoff and data traffic density.According to this scenario the data and traffic management signaling information is being routed to an anchor entity by using Centralized Mobility management solutions.However this will affect the reliability of this solution like ( single node failure) ,scalability and perofmance of the network.Based on these findings IETF DMM working group has proposed an innovative archeitecture named DMM [2] to resolve these problems[3].

It is important to notice here that , the requirement for DMM solution is to further broaden and reprocess IETF standard protocols,however they are more robust and less error prone[4]. So it is obvious that most of DMM solutions are fabricated based on the current CMM IETF protocols.Moreover DMM Solution as compared to CMM solution have been divivded on the basis of involvement of MU in IP mobility process and not on the basis of host based or network based .It has been studied vigoruisly that previous work has expanded MIPv6 standerd as host based DMM Solution[5] and have also expanded PMIPv6 protocol as network-based solutions [6-7].As its clear from the discussion that network based DMM had better through put then host based DMM[8].

In general, privacy mechanisms are used to provide partial disclosure of the user’s data. In LBS context, revealing the user’s location partially is required to provide the service for the user with an accepted level of utility. In reality, the required utility level varies based on the nature of each LBS service. For example, specifying the user’s city is enough to get valuable information from a weather forecasting service. On the other hand, the utility of a POI retrieval service degrades sharply if the reported location is several hundred of meters away from the user’s real location. Therefore, the user should have the ability to adjust the location privacy level for each LBS to get the required utility level. Many mechanisms have been proposed to preserve the LBS user's location privacy. The goal of these mechanisms is to blur or reduce the resolution of the user's location sent to the LBS server effectively. In other words, they aim to make the tradeoff between the location privacy and the service utility with an acceptable level of computation and communication overhead. While said mechanisms have had demonstrable effectiveness with snapshot queries, the shortcoming of supporting continuous queries is their main drawback. These mechanisms can be classified into k-anonymity, mix zones, cryptographybased, and obfuscation

Intelligent transpostation system network based DMM has meet several challenges.How ever during vehicle movement , which in turn will be responsible for high packet loss.The reason behind this is network based DMM which starts the handover procedure promptly as soon as the attachment of MU with the Mix-zone (MZ) has been completed.In this paper , we have given a network based DDMM technique which is going to extend the process of Fast handover for proxy Mobile IPv6(PFMIPv6) protocol which is going to propose Fast handover for DDMM .However PFMIPv6 protocol is being standaddized in [9] to lessen the handover latency , packet loss and session recovery of PMIPv6.It is also obvious from the discussion that FDMM emloys revised functions and messages from its normative PFMIPv6.However FDMM defines a two dimensional tunnel among serving mix zone and mix zones server to tunnel MU's packet.So this will help MU to send packet as soon as it is being connected to mix zones server and receive packet vice versa.According to our construct , we have defined two operation modes : one is called predictive (pre-FDMM) mode and another is reactive (re-FDMMM) mode.By performing above activities we scientifically analyse the throughput of FDMM by considering the handover latency , session recovery , packet loss and signaling cost with respect to DDMM.

**CENTRALISED MOBILITYMANAGEMENT**

As it has been known so for that , recent mobility management system techniques depend upon centralized and hierechical network architecture that is going to implement mobility anchor to route traffic from current terminal location.During this Home agent will manage the mobile nodes (MN) bind together among their home and address and tunnel traffic towards latest MN location.It is also considered that extension and optimization such as hurry handoff for Mobile IPv6(FMIP)[3], Hierarchical mobile IPv6 (HMIP)[4] and proxy mobile IPv6(PMIP) [5] is going to save the centralized and hiereachical procedure . There are certain option for managing global mobility in heterogenous architecture are also been considered , which are firmly based upon centralized functions , mostly depending upon MIP like procedures.lets consider ,[6] which is going to define multiple mobility solutions before giving mobility management structure by integrating different access networks where MIP is used.Moreover certain other protocols which are going to support network assisted techniques in which handover decision and triggers might be given at multiple levels in the network , where as the execution is also depends upon the usage of MIP. So this might be defined with the help of the concept named [7] micro mobility support in WLAN access networks and also in [8][9] for macro mobility in multi access system contexts.

**DISTRIBUTED AND DYNAMIC DISTRIBUTED MOBILITY APPROACH**

The main objective is to disseminate mobility traffic management, supporting multiple categories of mobility which might be dynamic user traffic anchoring in access network nodes. According to our construct as defined in fig -1 , where multiple three functanalities are being considered:

-according to this MN might be mobile terminal – The AN will be a "base station" that is going to provide the link layer and network layer connectivity to the MN as well as IPv6 address dissmentaion. According to eNodeB in E-UTRA [10], IP routing and tunneling opportunities are being provided as the same functional node as the base station one.

- During this , the AG would be standered border router , that might not be supporting particular mobility function ; it is very important to notice that the core architecture shown in fig 1 might not make any consideration based upon the defined topology.how ever it is worth to be noted that functional interfaces among relative access nodes may not require direct physical connection which is not a ring based topologies. All these gateways , access nodes and mobile nodes may taken advanatage from IPv6 standard [11] that might confirm allocation of multiple IPv6 address to the considered MN.Moreover when it is being connected with AN , the MN will receive an IPv6 address , that is being granted from the prefix defined by the AN.So that , the MN can start and manage data routing session by considering this address in a formal way , however it will be connected with the same AN.So forth it transferred to the new AN , then MN will get a new IPv6 address from the new AN and might be using it for the new session that would be created later on.however it is also been noticed that as long as the current session is being started the previous session are not going to be closed, so the related trffic might be forwarded among the last one and the new ANs. So we might be considering the useage of a simple tunneling technique to forward and compose traffic directly among ANs , as defined by GRE(Generic Routing encapsulation) [12].

The main contributions of our work are as follows:

1) This work provides real-implementation results for significant parts of the handover process which cannot be obtained through simulation. We consider that simulators though useful to some extent from an analytical point of view, either introduce unnecessary uncertainty into the road networks, or strictly specify significant parameters. Therefore, there is always a margin of error in simulation results. In addition, items which are not easily measured, are deduced analytically.
2) We use predictive handover by using an accurate prediction method is being proposed to improve mobile IP performance for Multiple mix zones over road networks. Our proposed method efficiently combines vehicle movement with probability analysis model to conduct well-informed predictions
3) We present a new handoff lantency procedure to obtain an optimal handoff opportunity to avoid untimely and unstable handover and unnecessary bicasting when the users comes to Mix zoens, which does not only depends upon the entering the mix zones but also on the exiting the mix zone .
4) We compare our DDMM method with re-FDMM , pre-FDMM and with other previous methods [35,36,37]. Our method reduce packet loss , handover latency and network scale parameters and has produced outpstanding results as compared with previous methods[35,36,37].

## 2. Related Work

In the current situation by considering the extension and adoption of current CMM proptocol that is to support DMM might help backword conformance and help the architecture of the network for the migration. Recent analysis [10-13] has revealed that network based CMM protocols are better then the host based CMM protocols.Therfore it is better to support DMM rather then host based protocol[8]. In the later section we shall explain the core solutions that might be defined for network based DMM.

According to [14] , the researcher has given an idea which is based upon very early partially distribuited proposals for network based DMM, that is being known as Dynamic Mobility Anchoring (DMA).However the core functions of mobility management are being transferred to a new entity called MAR that might have Mobile access gateway functanalities as well.The said technique depends upon the Database DB to place recent mobility sessions, and to give network extensions to MUs and their relevant MARs.So during the process of handover , a new entity of Mar is going to access the IP address information of the anchored MRs from DB.So that the latest Mar would be proceed with the binding update procedure.Nevertheless it has been noticed that the cost of signaling and as well the hand over latency is high when compared with others [15].

**A. Mobility Management for V2I Communications**

According to the recent study it has been seen that mobility management in cellular systems might be developed to acquire to multiple mobility scenarios.We have a example of 3GPP architecture , which might get provide multiple mobility procedures among 3GPP radio access networks or it might be in between different 3GPP radio access networks , or we may say that with non 3GPP based access networks.when we are going to consider , a new access network which is being known as the “Evolved UTRAN” or LTE (Long term evaluation) is being introduced , which will be considered with new packet core network objects , the serving gateways and PDN gateways[1][2].

**B. Towards new dynamic distributed Mobility schemes?**

As it has been discussed earlier , IP based and cellular based mobility techniques might uses centralized approach for tunneling to execute the the data path to MN’s location. There fore it is also been concluded that , data path might get authenticated by using multiple encapsulation levels , however MN is stationary . It is also depicted that , during handover take place , tunnel updates in the core network might influence the throughput by causing delays and packet loss. It is also been observed that optimization procedure predict direct forwarding of data traffic among near by access nodes, at the time of handover procedure, by adding encapsulation function and out of sequence packet delivery risk.However , it is assumed that both LTE and FMIPv6 handover schemes might be using some short interval tunnel among the older and new access nodes (eNodeB in LTE and AR in FMIPv6) to forward packets until the procedure of handover get completed.Moreover it is also been noted that , centralized encapsulation mechanism and the path management raises issues i.e network congestion and single point of failure , however a new distribuited approach might be given for mobility management scenario, this can be usefull either to intra-technology or inter technology mobility.[48-55]Therefore , a distibuited approach has been proposed for 3G SAE core network in [8] , in which MIP HA function has been distibuited among mobility agents and using DHT (distribuited hash Table ) to retaing binding caches.

It is also been considered that the mobility path must be managed and connected as near as to MN , hence by not considering the core network functions,henceforth we given in [9] a novel distibuited and dynamic mobility scheme , that is being named as DMA. This techniques is being refered as rest of this paper.

In [16] author has given partially distribuited technique which is based on PMIPv6 , which is being known as D-PMIPv6.According to this , mobility Anchor (LMA) have been divided into two separate entities: 1) Control plane

LMA(CLMA) , 2) Data plane LMA (DLMA). The first one is going to manage the the binding registration and going to elect DLMA for each MU , however the second approach might forward the data traffic.It is also been noted that MAG will manage the mobility related signaling message on behalf of MU and follow its movement as it has been done in PMIPv6.Soforth the presented technique may reduce the overhead on LMA, moreover it is going to continue the tunnel connectivity and hierarchy architecture among DLMA and MAG during a complete data session.So it is therefroe concluded that it might not fulfill the requirement of flat architecture. In [17] another approach named H-DMM has been presented as a partial distribuited approach.In this approach author took the benefit of CMM and DDMM solutions by smartly joining DDMM and PMIPv6 protocol. This protocol might get applied based on flow charaecterstic and number of active prefixes.According to this approcach , MU will select two different prefixes , first one from report server and the second one from LMA, however the first one is being changed with every handover , while the second one remain un changed.It is determined from the discussion that , author was unable to identify , how to regulate and select the appropriate protocol to get executed.

In [18] another hybrid centralized-distributed DMM technique has been given , according to this approach , two mobility anchors were proposed , 1) Access mobility Anchor (AMA) , which will perform the functanality of MAG and have been stretched to delegate a prefix to MU and 2) LMA , which execute the functanality of PMIPv6's LMA and perform the operation of creating tunnel with AMAs on request. It is therefore determined to establish tunnel for an MU by calculating session to mobility ratio utilized at AMAs , moreover it is also determined that tunnel creation will remain betweem AMAs until the session to mobility ratio is lower or equal to certain threshold.On the other hand , the tunnel will be created by MU's AMA for the MU with LMA, but the given scheme certain shortfalls , i.e there is a drawback of moving among LMA's during the mobility session,that is not suggested by IETF[19,20].Aprat from that , there might be the possibility of service intuerption because of updation of new anchor location of the prefix used by the current flows.

Another approache named fully DDMM is been given in [6] by delegating a network prefix to PMIPv6 domain by using different sub-network prefix for different MAG.In this approach functanalites of LMA have been disseminated among all MAGs , in which they would like to advertise network prefix and it is also been noted that each MU would determine its information by using statefull ip addressing, but there is a drawaback in this approach , it requires very solid control of IP addressing. According to [21] , very robust distribuited approach which is based upon PMIPv6 and IEEE 802.21 have been propsed. In this approach deployment of MIH function (MIHF) have been discussed and to update mix zones with reported server and mix zone server as handover occur.It has been concluded from the results that F-DMM has reduces the handover latency and also packet loss as compared with the P-DMM and CMM. However this approach lacking the information regarding how MIIS will get the information of near by networks.It is also been noted that , F-DMM would start pre-handover process as soon as the link condition becomes bad , so that the MU might not be connected with reported server send and receive BU messages.

Another approach has been introduced in [22] , according to this partially distribuited solution and the extension to fully distribuited solution has been proposed. This approach is one of the robust proposal for IETF network based DMM that would get published further in [15]. According to this , LMA functanlities have been distributed among all access routers , that are being named as MZ . Moreover DDMM will employ the idea of flatter system , in which the RSU entity is going to be fixed dynamically very much near to the MU. In this scenario two main ideas have been floated to construct DDMM; 1) partially dynamic distribuited technique 2) fully dynamic  distribuited technique .In the first technique , LMA's information plane is only going to be disseminated among mix zone , however the control plane will remain centralized towards reported server .According to RS approach , it will maintain mobility session controlled by exchanging proxy binding update (PBU) and Proxy Binding acknowledgment message with MZ.According to the definition MZ will going to manage local binding cache entry (BCE) of the MUs , that are being connected to it.Moreover every MZ has a different set of global IPv6 prefixes that are being known as local Network prefixes(LNP),that might be allocated to MU upon attachment.According to fully distribuition approach, data and control planes will be fully distribuited among MZ. It is also been concluded that the handover latency and packet loss of DDMM might considered to be not much different from PMIPv6[23].

**C. Mix zones**

Mix zone is location privacy principle for services which requires identifying the service user using a verified pseudonym [18] [19]. In reality, each user uses a pool of verified pseudonyms to preserve the location privacy. Thus, each query contains the user's pseudonym and his/her accurate location. Mix zone is defined as an area where users are required to stop sending the queries to the LBS server inside it. Accordingly, when multiple users enter a mix zone at the same time interval and stop sending their locations, their identities are mixed inside that zone. Furthermore, each user uses a new pseudonym when he/she leaves the mix zone. Therefore, the LBS server cannot link the old pseudonym of a user with the new one. As a result, the user's whole trajectory cannot be constructed. Similarly, exchanging the users' current and future pseudonyms and reusing them was proposed to preserve the location privacy [20]. In [21], pseudonym exchange schemes were proposed, which consider the user's speed and direction, and the pseudonym's age to reduce the exchange overhead. A centralized architecture is proposed in [22] for optimizing the mix zones placement using the mixing load sub-graph. A graph abstraction algorithm is also introduced to increase the anonymity level. In [23], a method for dynamic pseudonym changes is introduced. The method depends on the velocity and movement trends of vehicles to enable them changing their pseudonyms dynamically inside and outside the mix zones. An urban-adapted strategy for changing the pseudonym is proposed in [24]. The strategy supports the pseudonyms changing and exchanging inside the constructed mix zones.

Mix zone mechanisms suffer from two main problems. First, constructing and placing effective mix zones require extensive analysis of the road network characteristics and the motion patterns of the users. Such analysis requires extensive effort, and needs to be done periodically as a result of changes in the road network structure (i.e. opening or closing roads) and changes in the motion patterns of the users. The second problem is blocking the use of the LBS inside the mix zones, which may affect the service usage negatively.

**D.LIMITATIONS OF CENTRALIZED MIP APPROACH AND REQUIREMENT OF DYNAMIC DISTRIBUTED MOBILITY MANAGEMENT**

In this section we will discuss advantages of DDMM over MIP which is to avoid non optimal route for MIP , therefore the idea of DDMM is taking up numerious possibilities.Moreover the problems related with centralized approach might be solved at the same instant.Therfore the issues with centralized mobility management construct would be explained in [9].The problems might be 1) non optimized routes selection , related with content delivery network (CDN) servers , that is being displaced near to the user; 2) non- optimality in emerging network architecture ; 3) less scalability for centralized route and mobility context maintance ; 4) central point of failure and attack ; and last but not the least 5) non optimal usage of resources that would support mobile nodes which do not need mobility support.According to DDMM approach , mobility support is being available at different locations and it will always be near to the node, therefore the nodes might execute handover with session connectivity without routing through centralized anchor node in Data plane.However if control planes are being considered then the mobility management functanalities and other controls might be centralized.So by this way mobility management might be employed in a distribuited way so that the non optimal route in problem (1) can be vanished. It is also been concluded that mobility anchor that is being near to access network to which the mobile node is connected to ,therefore by giving solution to the flattend network and CDN network might be solving problem (2), another approach which is being known as distribuited deployment that is more robust then centralized approach might be addressing problem (3) , and in the last it is going to overcome the problem of centrliazed point of failure and attack addressing problem (4)

## 3. Our Proposed Method

### 3.1 WHY DYNAMIC DISTRIBUTED MOBILITY MANAGEMENT?

According to [4] , it is very much obvious from the previous discussion that Quota of generation real time traffic has been dramatically increased and delivered to smartphone and tablets rather then traditional PCs, This process might get emerging not because of mobile network architecture but need to change current CMM artechticure. It is also be noted that DDMM technique among flat mobile network architecture has been evolved as a counter act mobility management that would remedies the latest CMM limitations.It has been depicted in fig 1 that CMM would need single handed mobility anchor , that woulb be HA at MIPv6 and LMA at PMIPv6 , which is being required to create a session continuity when MN would like to move among hetrogenous networks.It is also been clear that mobility anchor would required to manage normal routing with tunneling support for the registered MNs.It is also required for mobility anchors to manage all mobility contexts and routing procedures for the MNs.However by creating an environment in which mobility anchors may follow centralized architecture becomes hard to extend the environment in which more numbers of MNs are required [5]. Moreover it is also clear from the discussion that as mobility anchor are disseminated via DDMM , therefore high scalability and availability for mobile internet service might be provided and at the same time removing the threat of sinle point of failure.Therfore , it is evident from the discussion that no single handed anchor would be there to support mobility is being deployed in DDMM however they are being distribuited at multiple road side units[56-60].

So , it is also concluded that load to support mobility in such networks are being processed in a distribuited way, however the chances of failure of a mobility anchor among other is very less.Moreover by using DDMM approach will also help to resolve the issue associated with centralized performance bottleneck and also execute traffic that is related to MNs in a distribuited manner. As discussed previously , in CMM architecture only single handed mobility anchor is being used in tunneling for traffic forwarding among all registered MNs [3,6,7].Thus it is obvious that single handed mobility anchor would considered as a centralized performance bottleneck , that would cause poor network performance by enhancing end to end packet transmission delay.By considering DDMM , it is very much obvious that no centralized mobility anchor would be required so hence forth bottleneck issue might be avoided.Moreover , it is also clear that MN would be served that would be close by mobility anchor and situated at the first hop router so forth the end to end packet transmission performance might get enhanced and that would remove the threat of sub optimal routing issue of CMM [3,6,7].Moreover it is also been noted that it is not required for every MN that might require mobility support during their handover. Its all based upon the type of execution at different MNs.It is also been concluded that CMM is not specially designed to manage efficiently the resources for MNs that does not require mobility support.However in DDMM , the procedure has been designed in such a way that , the upperlayer would only be provided transperncy when they actually neeed it , and at the same time procedure such as prefix provisioning , address configuration and signaling message to update location would get minimized.

### 3.2 DYANMIC DISTRIBUTED MOBILITY MANAGEMENT

According to DDMM approach , it will use existing IP mobility protocol that might be MIPv6 and PMIPv6 for the emerging flat IPv6 mobile network architecture.The main functanalities of this to disseminate and confine the function of mobility support at the part of AR level, however the other network would e unaware of the mobility and their support.It can be explained in such a way that , as the funcatnaility of mobility anchoring is being ditribuited at every AR , and at the same time DDMM only be activated when mobility support would be needed, that might be when MN would pass through an IP handover.As we get it comare to MIPv6/PMIPv6 in which the data traffic is being connected with the same entity that might be HA/LMA , however dynamically mobility management would be changing the anchor for a complete new session . It is obvious from the discussion that  the new session in DMM is being connected with the latest mobility anchor (which is being deployed at AR ) and being executed by using the current IPv6 address at the same time.It is only be possible because IPv6 allow MN to use different IP addresses at the same time.Henceforth the data traffic would be flowing down optimaly among MN and CN without using the concept of tunneling , until and unless MN will passes through an IP handover. Moreover it is stated that if MN passes through more then one handover before ending up the session , then it is obvious that data traffic of the said session will be forwarded via mobility Anchor . So , we can say that DDMM is going to optimize the routing path for a very high amount of percentage of the data traffic [10]. It is also been fact that at a particular instance more then 60 % of the said users in an operational network are non mobile.Apart from that DMM is going to reduce the tunneling overhead and global network signaling load , the reason is that the mobility anchor are very much closer to the MN . Moreover , the issue of single point of failure in MIPv6/PMIPv6 are also expected to be removed by usin DDMM approach , because of the binding management is being disseminated at the ARs level rather then being managed at the same entity.

### 3.3 MIPv6-based DDMM for Global/Local Mobility Support

According to this approach [9] , it will need to support global mobility as well as local mobility. It is very much clear form the above discussion that the MN will roll around between multiple access networks and operational domains.However it is not sure that the MN will always be connected with an HA ; it is also been deduced that MN may connect to some old AR that might not act as mobility anchor.Moreover it is also possible that the MN will connect to some MIPv6 network . This will make sure to no any modification in HA functanalities in the MIPv6 based DMM technique in order to be compatible with those MIPv6 networks. Therfore , MIPv6 based DMM has to explain the DMM basic modalities and must take into consideration multiple scenarios and condition that are being discussed above. So we have distinguish among two different techniques as defined below.

### 3.4 DDMM-mix zone Operation

According to our proposed technique FDMM that relied upon IETF network based DDMM , it is been concluded that it is still in the novice stage of standardization [15]. In the next sub section we will present the detail of network based DDMM operation , that would explain the initial registration and handover process.

#### 3.4.1Initial Registration

According to this once an MU be the part of subnet , it will forward RS message to the particular mix zone in figure 1). After this , mix zone would allocate LNP (pref1::MU1/64) to the MU and will create local BCE for the MU that might contain MU-ID , LNP and other important parameters.After this , it might send PBU to the LBS server to start a new mobility connection and create its BCE. As the CMD would respond with PBA , the mix zone might forward a route advertisement (RA) message to the MU , that might contain assigned LNP used to configure an IP address. Henceforth Mix zone is being known as reported server as long as MU is being connected.

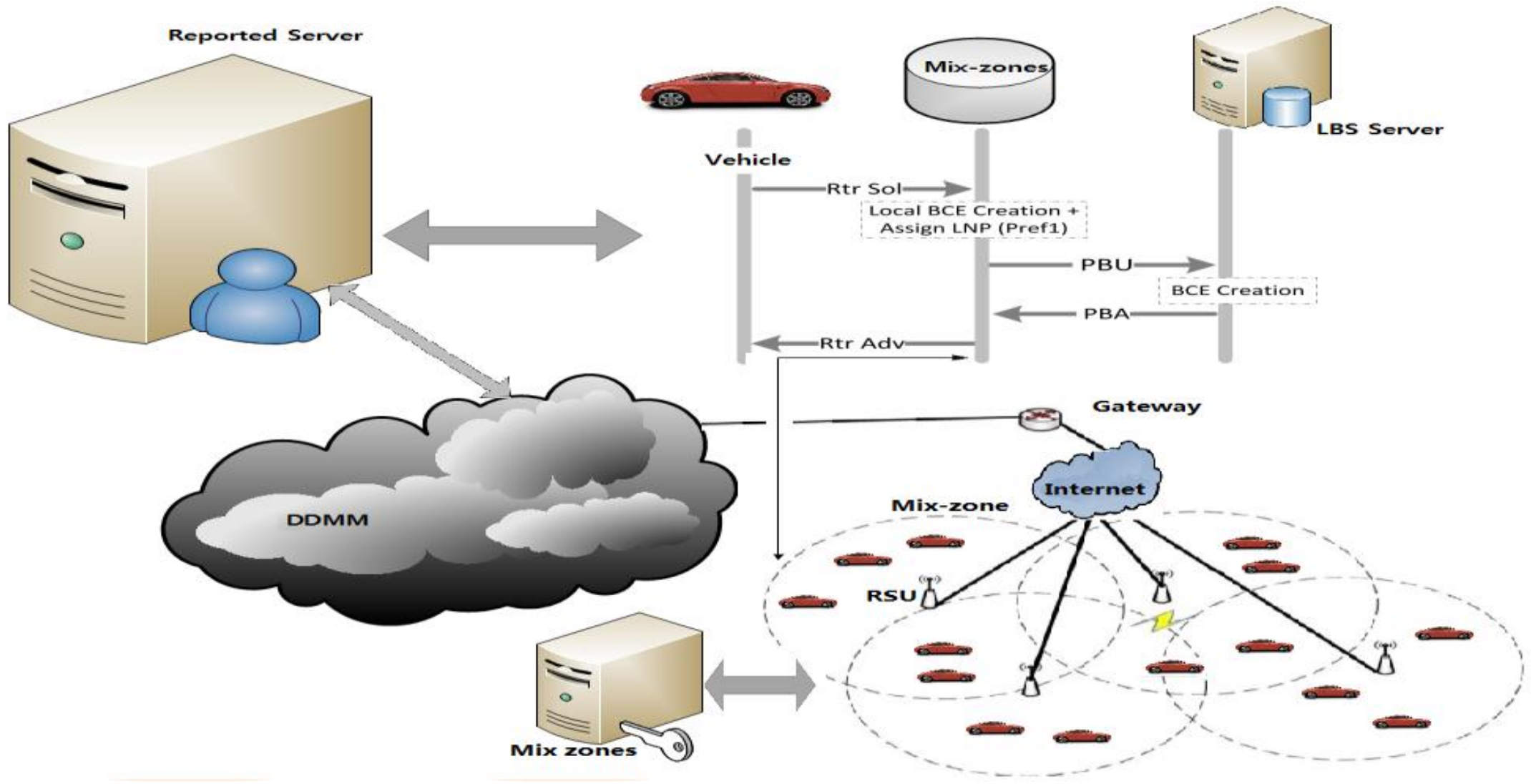

Figure 1: DDMM Initial Registration Operation [15].

### 3.5 Handover Management

According to this construct handover process wil begin as soon as the MU will move to different mix zone here would be mix zones as shown in figure 2. Moreover MU might send router solicitation message to the mix zone which is going to allocate another LNP for the MU by selecting from its set of prefixes and store temporal BCE. After that ,mix zone would send request to LBS server by the help of PBU message to get register the MU.As soon as the PBU will be received , the CMD is going to update the BCE by replacing the reported server information field with mix zones information field to every mix zone information inside mobile user field. MoreoverL BS server would reply mix zone by using an extended PBA that might contain the MZ address and the associated pLNP. As soon as MZ would receive PBA , in the mean time RA message would be sent which is going to advertise the new LNP (nLNP) assigned to MU.After that , MU will be configuring a fresh IP address (Pref2::MU1/64). Moreover it is also been noted that a tunnel has been created between mix zone and road side unit to manage the traffic of previous LNPs (pLNPs, Pref1::/64 in our example).Apart from that LBS server might exchange PBU and PBA message with active mix zones(MZ's in our example) to let them notify regarding new MU position and how and where they would update their local BCE and adjust the tunnel towards mix zones.

Therfore it is been seen that the active IP flow by using pLNP would be RSU with mix zone and at the same time the new flow using nLNP might get forwarded by using mix zone sewrver to the MU without any special packet handling.As we know that DDMM technique mean to say dynamic mobility , moreover it is also been clear that this will only be happen if the previous IP flows might require session continuity for example type of application that must not have to change its IP address during its information flow.So in that case , MU might have multiple IP flows connected at different mix zones.

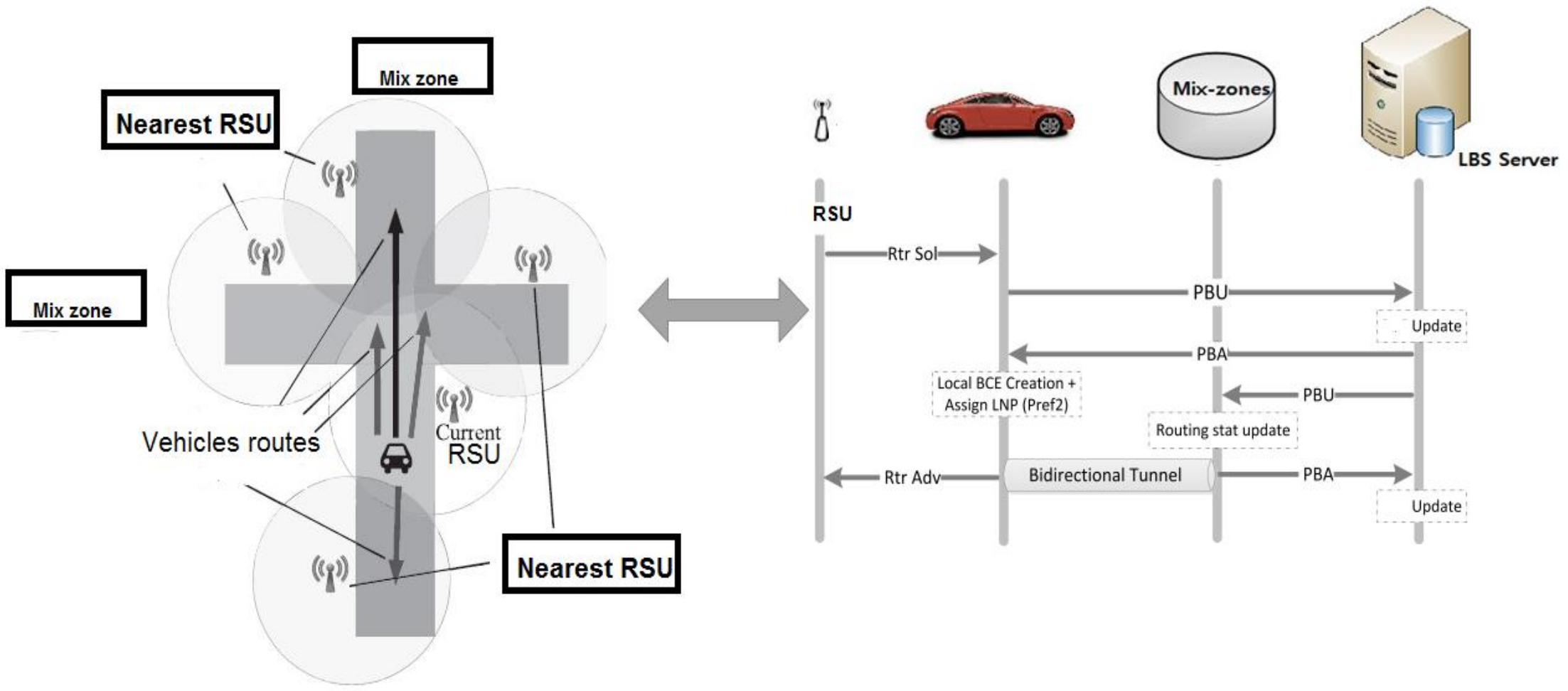

Figure 2: DMM Handover Management Operation [15].

## 4 Description of the Solution

According to the researcher the mix zone server inside DDMM is being considered for starting binding registration process for MU with LBS server.Therfore , it might be possible to enhance the handover process , if it is being communicated to mix zone regarding MU attachment, message format and its functions.However if we consider FDMM , the first one is called as reactive-FDMM and the second one is predictive-FDMM mode. Moreover FDMM would support both of them like fully and partial DDMM model defined in [15] while the LBS server might not participate in handover management process.In the following section we will explain FDMM message and mobility options and further explanation of the handover management operation.

### 4.1 FDMM Messages

So to execute mix zones server and to get associate with MU and to restore the previous session , the Hand over initiate and handover Acknowledge messages in [9] are going to be extended for the sake of context transfer , for that parameters values like MU-ID , MU's LNPS , reported server address , LBS server and Routing state table are being shifted from the reported server. Moreover we are going to explain few mobility options that might be appended in HI and Hack message.

A. **Handover Initiate Message (HI):** according to this , it might be received on mix zone with RSU coverage during pre-FDMM mode , however during re-FDMM mode , it will be delivered to reported server .Moreover the message type Distribuited (D) and target type (T) are being inserted to the message format that are being explained in PFMIPv6. So D flag might be used to distinguish the messages that are being discussed in [9], moreover its value must have to set to one in all new messages by using prposed extension.While the T flag is being used to differeniate if the receiver is mix zone with 0 value and if reported server with 1 value or reported sercer with the value 2.

B. **Handover Acknowledge Message (HAck):** According to this , it might be sent by mix zone or RSU acknowledging the receiving of HI message . It can be the same as explained in HI message , therefore D and T flags are explained in Hack message.

So it is also explained that the mobility option defined in HI and Hack are little changed might be depending upon the given information .So the option for mobility are as follows:

## 4.2 Handover Management Operation

### 4.2.1Predictive mode

According to this , handover mechanism came into execution in pre-FDMM mode as soon as the RSS related to MU and anchored with current Access Network have been decreased to less then a certain threshold value i.e $S_{th}$.Moreover the MU will find a new AN(nAN) (i.e by doing scan of its near by ANs in which the nAN will be that from which the MU will acquire strongest signal).So the process starts when MU dispatch L2 report message that might contain its ID and the nAN's ID , towards the RSu ( here in this case in this example as shown in figure 3).As soon as the report message would be received , RSU would drives the near area of mix zone from (AN ID ,mix zone area info) information in the report message,very much similar to [25].However it is also deduced that the layer 2 handover might occur if the MU will be moving around the same mix zone . However , if the MU is heading towards mix zone , then repoted would dispatch HI message with value zero of "T" flag and value one of "D" flag to themix zones server which contain the MU-ID,MU's LNP (i.e., Pref1::/64).While the "D" flag represent that it is distribuited extension however "T" flag represent that HI message is for mix zone. Therfore LBS server and MU LLA-IID option might also appended in HI message.

By acquiring the HI message , mix zone server would allocate MU's nLNP (Pref2::/64) and store it into temporal BCE at local database. Henceforth m would also refresh the routing state.However the LNP that might be allocated prior to this is now known as an MU's pNLP, and at the same time relative mix zone would perform as RSU for that LNP.Thefore , mix zone would respond with Hack message and will allocate LNP to the reported server to confirm that the updated location has upgraded successfully and the routing information has been updated.So two way path has been created among reported server and RSU, and the information that is for the MU is being dispatched from the reported server to the mix zone over that path. It is also been noted here that the traffic dispatched to any any MU's prefix (i.e multiple handovers) might be sent to the mix zones. Henceforth mix zones would initiate buffering once it will receive data traffic.

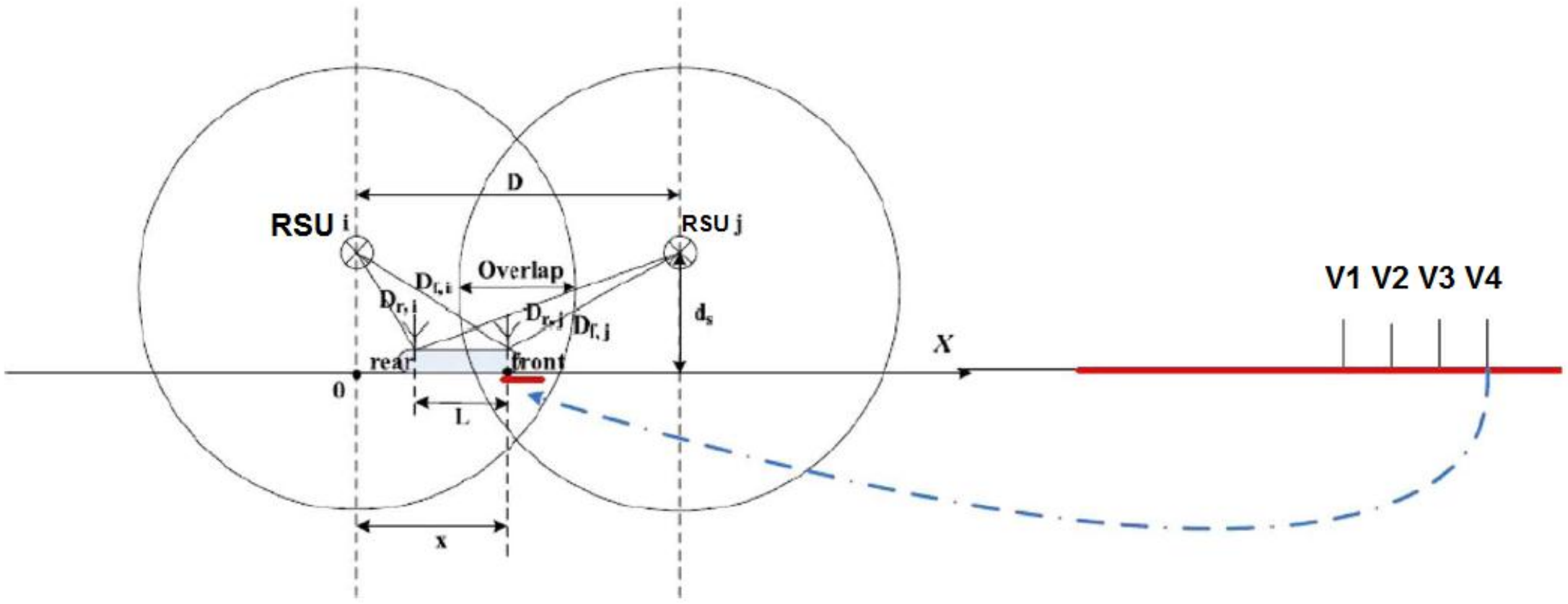

Figure 3: Predictive-FDMM

So as the handover procedure is ready over the network , MU might start performing handover to nAN. Reported server would send in handover command messages , and assign LNP to the MU , that also assign a new IP address (Pref2::/64).Apart from that , Pref1::/64 is also deprecated , by broadcasting it with a zero second lifetime.Therfore MU would create a physical connectivety with the nAN (e.g., radio channel assignment) , that in response initiate the link layer connection among the nAN and mix zone if it was not created previously.So by this way , flow of old IP using Pref1::/64 are in working condition and connected to RSU, while the latest flows might use Pref2::/64 , that might be the LNP routed by mix zone regardless of any special handing of packets for downlink and uplink as depicted in figure 3. So now the mix zones would dispatch the PBU to the RSUs with latest mix zones and LNP options. As soon as the PBU received and BC lookup , the LBS server would acquire previous entry for the MU and would upgrade th MU's cache by replacing the Address of S-RSU with Mix zone 2, and also append the last location (mix zone 1's address) into mz server list field in addition with the pLNP assigned .Moreover , LBS server would dispatch back the PBA to the mix zones .Hencforth , from now onward , the packets coming to/from the MU would pass through the mix zones 2 rather then mix zone1. So the miz zones2 act as mix zone server.

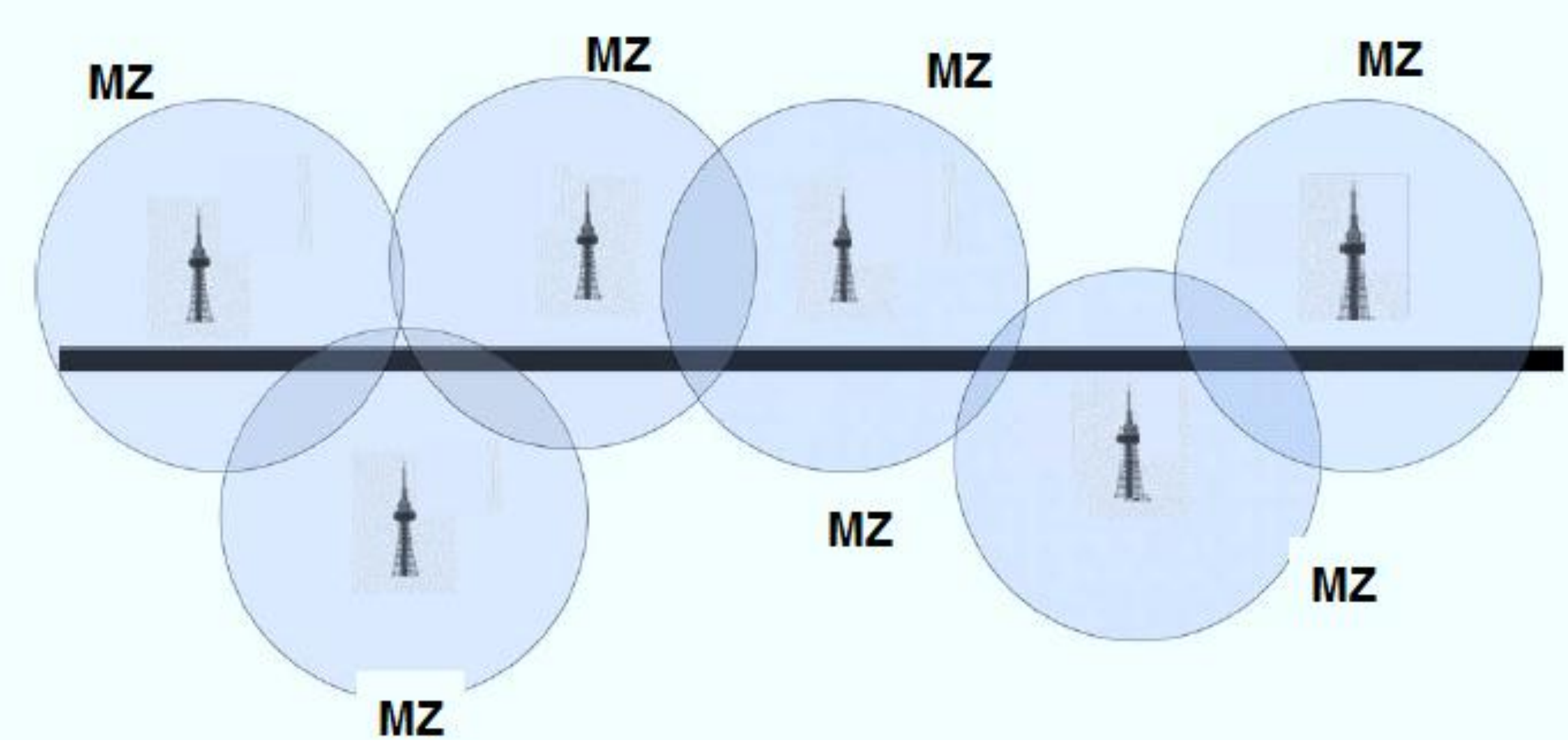

Figure 4: Predictive-FDMM

#### 4.2.2Reactive Mode

According to this mode, MU would go through a handover procedure by the sAN to the nAN. So the , MU would create a link with nAN , which execute the link among the nAN and mix zones server. The MU ID would be dispatched to the mix zones sever for the next procedure.Moreover the AN-ID over the previous connection , which might be given by the MU or nAN , might also be dispatched to the mix zones server to recognize the reported server over the new link.Therefore mix zoner server would dispatch the HI message to the reported server. However the HI message would have the "D" and "T" flags are being set and also the MU ID.Moreover the value of "T" flag is being set to two which might represent that the receiving end would be reported server.The context request option might contain and have request context information regarding MU to the reported server Moreover reported server might dispatch the Hack message back to the mix zoner server by setting flag “D” and “T”. Therfore Hack message contains the MU-ID , MU’s LNP, MU LL-ID , if and only if it is valid i.e non zero, and at the same time the LBS server address that would currently serving the MU.Moreover the requested context information by mix zoner server must be incorporated.Therfore a tw way tunnel has been created among reported server and mix zoner server , and the information sent for the MU is beinf dispatched from the reported server to the mix zoner server over this tunnel. Henceforth the remaing procedure and the data traffic flow will remain same as explained in predictive mode.

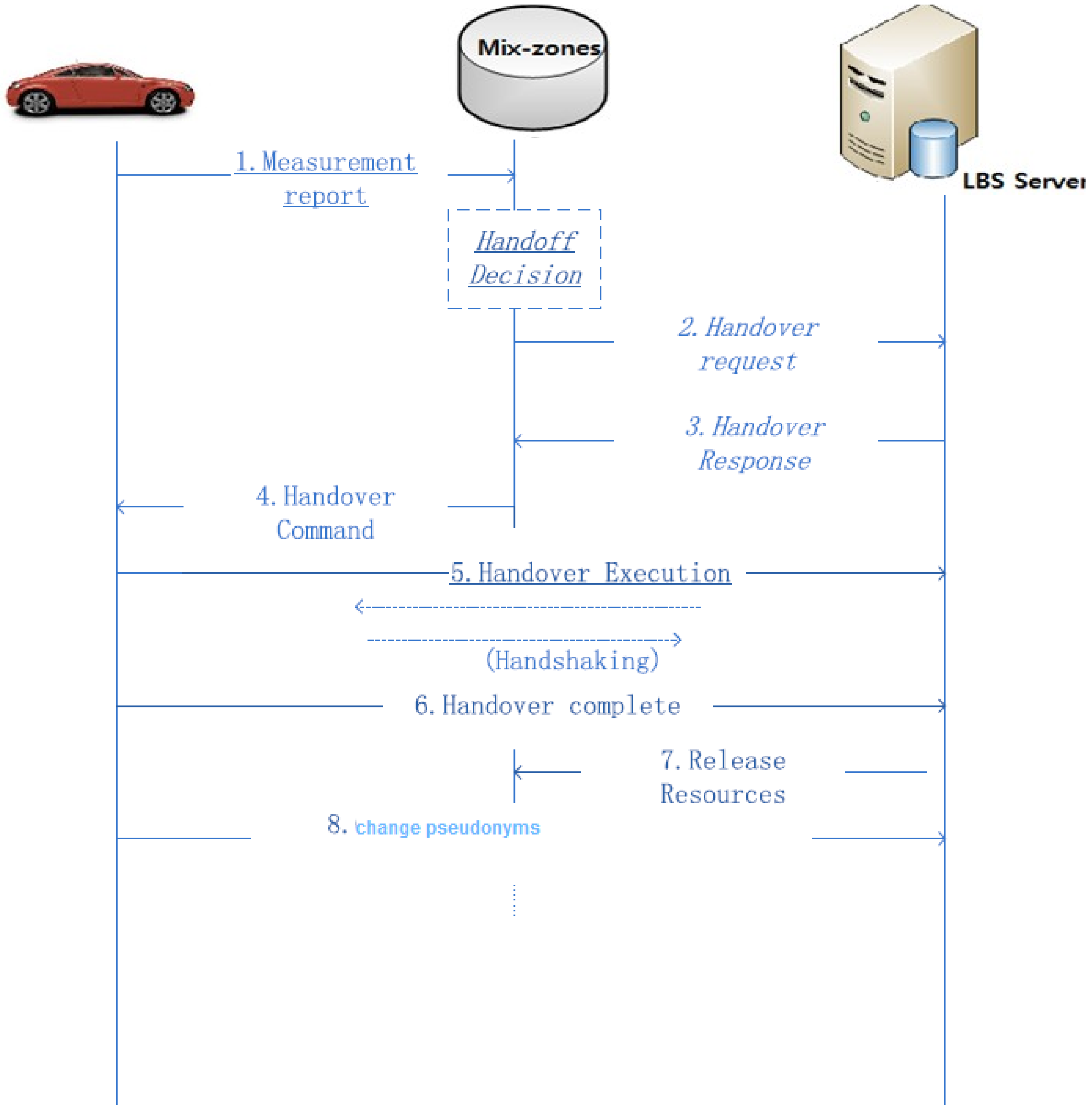

Figure 5: reactive-FDMM

If we consider next MU’s movement , the procedure would get repeated for reported server involved. However , as soon as the reported server dispatch Hack message towards the mix zones , it might be adding RSU mobility option that might contain the RSU address and the relative pLNP (refer figure 6). So , the mix zones server would dispatch HI message with one value of “T” flag towards RSU which coprises the MU-ID and latest mix zone option.However RSU would refresh their routing information by placing mix zone server as next hop count and respond with Hack as reply.

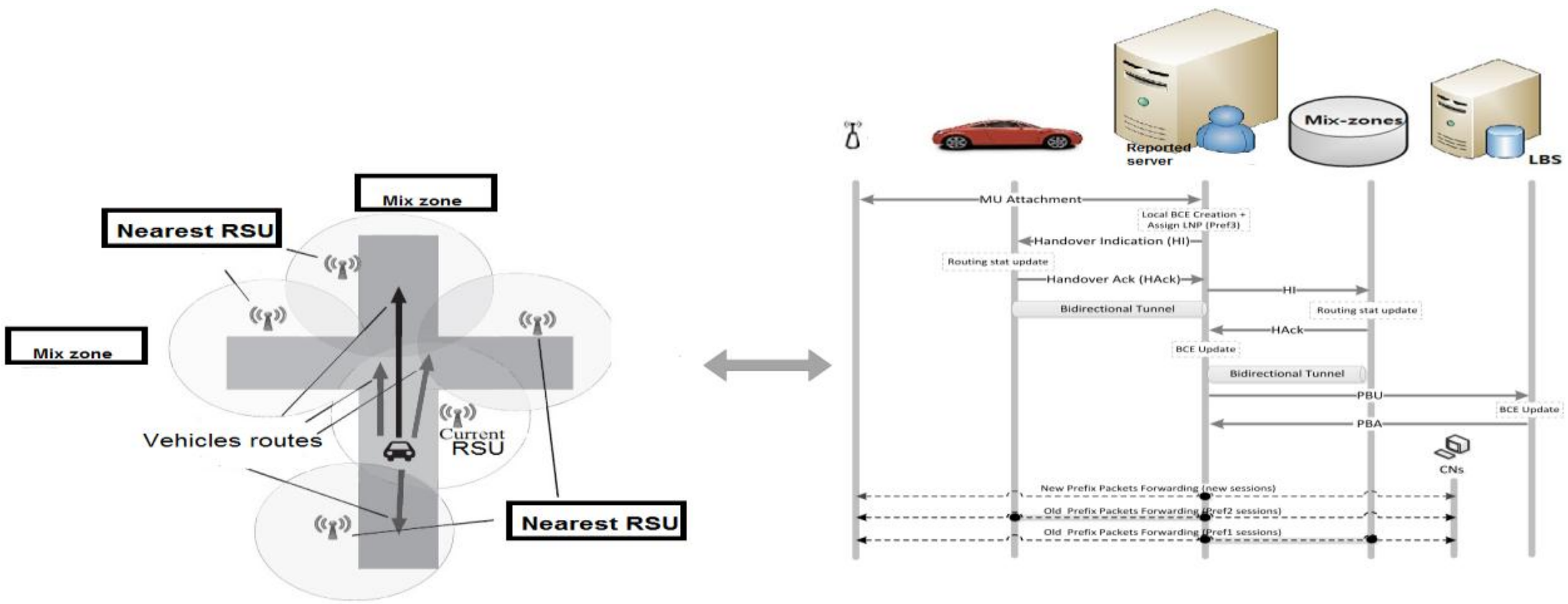

Figure 6: reactive-FDMM (re-FDMM)Handover Management Operation in Second Roaming

## 5 Performance Analysis

After having detailed analysis of the current solution we have designed an analytical model which will give us the idea regarding handover performance of FDMM as compared to DDMM. We have analysed different aspects .We have concluded that by doing analysis of these metrics will help us a lot to assess the handover performance of our given scheme on the basis of multiple system parametrs. Previously it can be seen that , this typ of analysis has got a lot of attention during different phases of cellular networks [26-29], and then this might get applied to the hand over and mobility management based IP-networks [30,31]. Notation used are described in Table 1.

Table I: List of Performance Parameters

| Symbol | Definition | Symbol | Definition |
|---|---|---|---|
| $SN$ | Number of mix zones(i.e. subnet). | $M$ | Number of mix zone entering/existing an environment. |
| $n$ | Number of mix zones in each row. | $r$ | Radius of a mix zone. |
| $X, Y$ | Horizontal and vertical area of analysis area, respectively. | $l_x, l_y$ | Overlapping space between two successive mix zones horizontally and vertically. |
| $a$ | Apothem of a mix zone | $N_x, N_y$ | Number of horizontal and vertical roads. |
| $S_x, S_y$ | The distance between two horizontal and vertical roads, respectively. | $h_{x,y}$ | The symmetric average number of hops between network entities x and y. |
| $\xi$ | Network scale. | $U_{max}$ | Maximum pause time. |
| $\bar{v}$ | Average speed of MU. | $\mu_{SN}$ | Subnet (i.e. mix zone) average crossing rate. |
| $T_{SN}$ | MU’s average residence time in a mix zone. | $T_{PR}$ | Active prefix lifetime. |
| $\bar{T}_{PR}$ | Mean value of $T_{PR}$. | $T_{PR}^{H}$ | Active prefix lifetime while the MU is attached to the inside mix zone(prefix used as LNP). |
| $T_{PR}^{F}$ | Active prefix lifetime while the MU is visiting a mix zone area (the prefix is used as pLNP). | $1/\lambda_{PR}^{F}$ | Mean value of $T_{PR}^{F}$. |
| $\bar{N}_{PR}$ | Average number of used active prefixes. | $\bar{N}_{pLNP}$ | Average number of active anchored prefixes. |
| $\alpha(H)$ | Probability that H handovers occur during the interval $T_{PR}^{F}$. | $E(L)$ | Expected epoch length. |
| $E(T)$ | Expected epoch time. | $E(P)$ | Expected pause time. |
| $L_2$ | Layer2 Handoff latency. | $L_{Auth}$ | Authentication latency. |
| $L_c$ | Average control packet size. | $L_d$ | Average data packet size. |
| $BW_w$ | The bandwidth of wireless link. | $BW$ | The bandwidth of wired link. |
| $l_w$ | Wireless propagation latency. | $l$ | Wire propagation latency. |
| $p_f$ | The probability that the wireless link fails. | $PC_x$ | Processing time by node x. |
| $C_{x,y}$ | The symmetric cost to transfer a packet between network nodes $x$ and $y$. | $d_{x,y}(p)$ | The symmetric delay of a packet of size $p$ sent from $x$ to $y$. |
| $\lambda$ | The sessions arrival mean rate to an MU. | | |

Following are the assumption that have been made in our analytical model (refer to Figure 7):

i.) It has been assumed that mostly every road network has been made of similar mix zone structure . Every mix zone is geometrically hexagonal having radius r of its circle area. If we consider a worst case scenario then it might be possible that each mix zone might have different IP subnet (*SN*).

ii.) It is been considered that at a distance D from the lat point of reported server , the value of RSS from reported server reduced below $S_{th}$. Moreover the relevant value of $S_{th}$, and the obtained value of Dis by the path loss model, $P_r(D) = P_r(D_0)\left(\frac{D_0}{D}\right)^e$ [32, 33], henceforth $P_r(D_0)$ is being the power acquired at a particular location having distance $D_0$ and $e$ might be the path loss exponent. However the value of $e$ might get vary and depends upon the environment. [33].

iii.) It is also been considered that MU would be acquiring RSS strength from mix zones at a specific distance D , however ,if MU comes out of the area H,moreover it might still feel some RSS strength from reported server.It is also been noted that MU would receive signal from reported if the $RSS > S_{min}$.

iv.) Handove initiate process would get started by MU to move towards mix zones server as soon as the RSS value coming from reported server shorten down below $S_{th}$.

v.) Moreover the process of buffering have been initiated by mix zones server as soon as it receive data traffic from reported server .At the same time we may consider this process in a downstream direction.

vi.) As we all know that the objects will move away from parent mix zone (that is the first mix zone it is attached to ). That would be increasing the mix zone-mix zone hope count (i.e.,$h_{MZ-MZ}$) among reported server and mix zones server for each and every time the hand over process will occur.On the other hand , we may assume that $h_{LBS-MZ}$ distance is the same for all mix zones,and the difference between them is very negligible. We have assumed this assumption as the worst case for our scenario , the reason is the signaling cost would get increase in an exponential way .

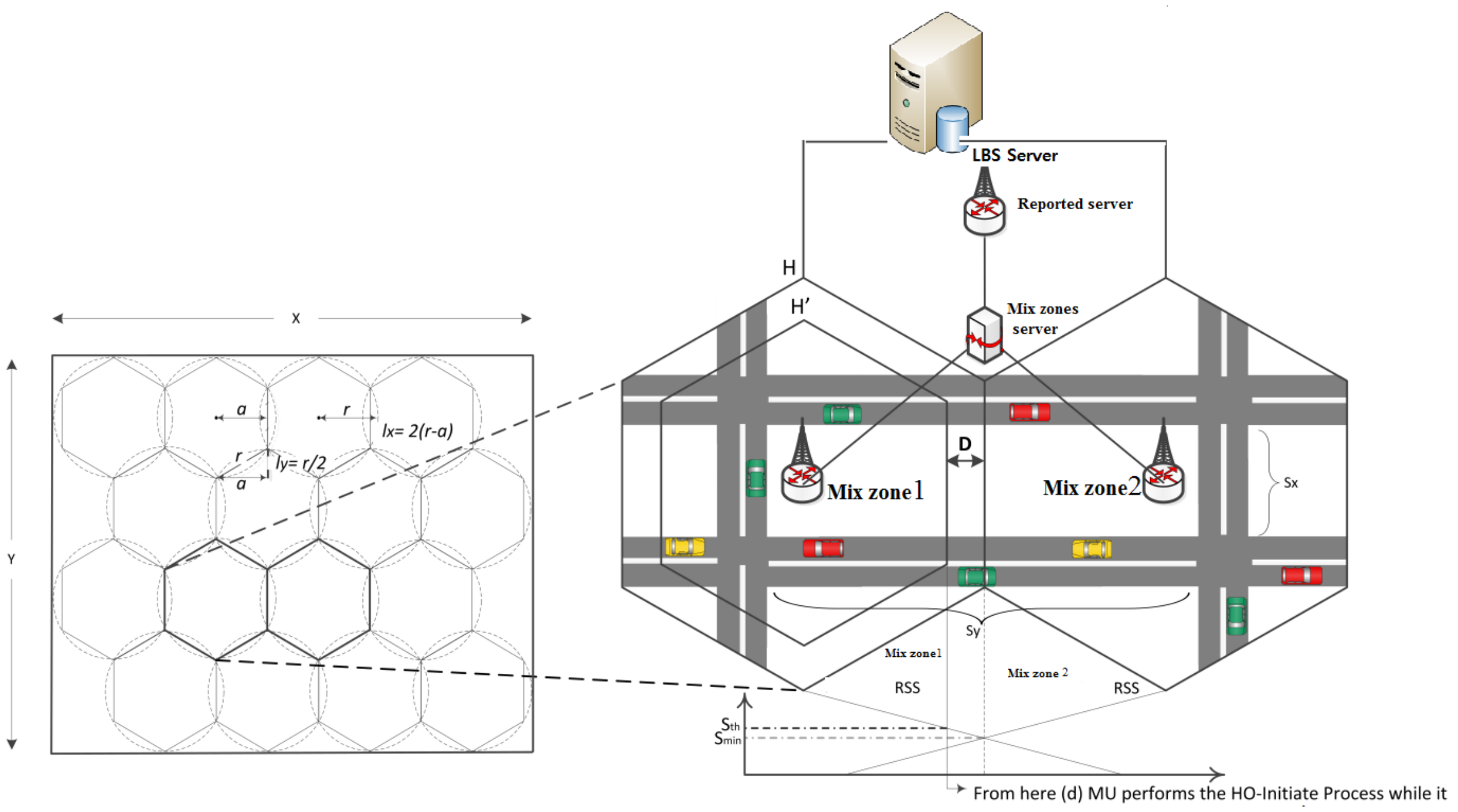

Figure7: System Model for Handover Analysis

### 5.1 Analytical Models

According to our scenario we are assuming rectangular area having dimension *X**Y* , as depicted in fig 7 . In this mix zone the number of subnets can be defined by this formula *n**m*. However the coverage area for radio transmission for eachmix zone will be hexagonal having radius r of a circle. Therfore two neighbouring mix zones might get overlapped at length of $l_x$ and $l_y$ with their horizontal and vertical diameters, individually. Then,

$$X = (2r * m) - ((m-1) * l_x) \tag{1}$$

$$Y = (2r * n) - ((n-1) * l_y) \tag{2}$$

Where

$l_x = 2(r-a) = 2\left(r - \left(r\frac{\sqrt{3}}{2}\right)\right) = r(2-\sqrt{3})$and$l_y = \sqrt{r^2 - a^2} = \sqrt{r^2(1-\frac{3}{4})} = \frac{r}{2}$

Thereafter , *SN* might be given by

$$SN = m * n = \left(\frac{X - 2r - r\sqrt{3}}{r\sqrt{3}}\right) * \left(\frac{2Y - r}{3r}\right) \tag{3}$$

According to the scenario we assume that the MU and the RS would be in the same domain area.As RS would be reported server as depicted in fig 7 and also represent the system architecture of our design.Moreover, $\xi$ ,is going to represent the ratio among the number of hop counts among two mix zones and the number of hop count between mix zone and LBS server.

$$\xi = h_{MZ,MZ}/h_{MZ,LBS} \tag{4}$$

where $h_{MZ,MZ} = \sqrt{SN}$.

After having the above discussion , it has been concluded that the number of hop counts between two adjacent neighbours mix zones must be less then that is among mix zone and LBS server. It has to be concluded that the network scale is $\xi \leq 1$.

However previous research has given the value of this ratio in between 0.2 and 0.5 [34–36]. According to our case , the default value might be $\xi = 0.5$. Therfore we shall discuss later that what could be the potential impact of this parameter on different cost.

To understand the concept of mobility modeling , we have chosen city section Mobility (CSM) model that is being introduced in [37] and [38]. This model is going to represent the behavioral movement that is being affected by certains environmental constraints.According to real time construct , vehicles might not move as freely as they want to because they have to obey some trffic regulations. When we are talking about CSM model it is going to represent a realistic movement pattern for the moving vehicles in a city [39], and its quantitative properties are being studied in [40].Thats why we have chosen this model for our mobiloity pattern. Acccording to CSM model the area is being represented by grid of streets that is going to formulate a specific section of the city. It is been noted that MU would start from a particular intersection of two streets , and further it will select a destination. Furthermore , as MU will reaches at particular destination , it will stopr for a some time and select another destination randomly and such process is being repeated. Moreover every cycle for MU to move from a source to a destination is being known as epoch[40].

As it is been very much obvious from figure 7 that , the roads are being lying parallel to axes . Nx and Ny represent the horizontal and vertical roads . It is also been shown that horizontal roads are being Sx distance apart and vertical roads are being Sy distance apart. Then, $N_x = \frac{X}{S_x} + 1$ and$N_y = \frac{Y}{S_y} + 1$. The anticipated epoch length is obtained as follows [40]:

$$E(L) = \frac{X(N_x + 1)}{3N_x} + \frac{Y(N_y + 1)}{3N_y} \tag{5}$$

Its also been consider $\bar{v}$ be an average velocity for a vehicle. Then, the estimated epoch time might get calculated as

$$E(T) = \frac{E(L)}{\bar{v}} \tag{6}$$

Moreover to consider safety at road intersection , there would a time which is being called as pause time , that would be between 0 to *Umax* to escape collisions. Therefore, the aniticipated pause time might be shown as

$$E(P) = \frac{U_{max}}{2} \tag{7}$$

By supposing that $2r = K_1 * S_x = K_2 * S_y$and $l_x = l_y = S_x/2 = S_y/2$, the estimated number of *SN*s passage can be acquired as in [39] and [40] as:

$$E(C) = \frac{mK_1(m+1)}{6N_x^2}(6N_x - 4mK_1 + K_1 + 3) + \frac{nK_2(n+1)}{6N_y^2}\left(6N_y - 4nK_2 + K_2 + 3\right) \tag{8}$$

As a final point, using (6), (7), and (8), the average residence time in an *SN* is expected as [36]

$$T_{SN} = \left(\frac{E(T) + 2E(P)}{E(C)}\right) \tag{9}$$

Then, we can acquire the crossing rate $\mu_{SN}$ by means of (9) as follows:

$$\mu_{SN} = \frac{1}{T_{SN}} \tag{10}$$

So it has been obvious that during the visit at each SN , vehicles in DDMM and FDMM might configure new address while the vehicle may use more then one IP address at the same time.As soon as one IP flow uses the address that might be derived from it then the prefix is considered to be active.Hereafter we have been using DDMM traffic model was used in [23].So it has been observed that number of active prefixes at a handover time might contain those , that is being configured from the LNP by reported server, and another one that would be configured from the pLNPs by RSU.So by considering DMM[15] , it is concluded that the prefix will be maintained atleast for the first handover , and then it might get expired based upon the user activity.So we can obtain the active prefix lifespan as follwos:

$$T_{PR} = T_{PR}^{H} + T_{PR}^{F} \tag{11}$$

In above equation ,$T_{PR}^{H} = T_{SN}$, and $T_{PR}^{F}$ represent random decay interval since the MU leaves the “home” mix zone unless the prefix expires. By representing $1/\lambda_{PR}^{F} \triangleq E[T_{PR}^{F}]$ , we might say

$$\bar{T}_{PR} = E[T_{PR}^{F}] = \frac{1}{\mu_{SN}} + \frac{1}{\lambda_{PR}^{F}} \tag{12}$$

So during the process of handover , it has been assumed that the average number of active prefixes,$\bar{N}_{PR}$, represented by one (the LNP) plus the average number of active pLNPs, $\bar{N}_{pLNP}$ :

$$\bar{N}_{PR} = 1 + \bar{N}_{pLNP} \tag{13}$$

$\bar{N}_{pLNP}$is the multiplication of the average number of handovers that might happen between the foreign prefix lifetime, that is in our case might be *g*= 1 prefix per handover.Therefore we have been refered to the probability that H handover might be possible during the time span of $T_{PR}^{F}$as $\propto (H)$. Moreover , it is concluded that pLNP will remain active for $E[\propto (H)]$ subnet residence intervals and from the above discussion , we got:

$$\bar{N}_{pLNP} = gE[\propto (H)] = E[\propto (H)] \tag{14}$$

In the last we received the probability $\propto (H)$ and its value might be:

$$\propto (H) = P_{PR}^{H} - P_{PR}^{K+1} = P_{PR}^{H}(1 - P_{PR}) \tag{15}$$

$$E[\propto (H)] = \frac{P_{PR}}{1 - P_{PR}} = \frac{\mu_{SN}}{\lambda_{PR}^{F}} \tag{16}$$

By combining (13), (14) and (16), finally we get:

$$\bar{N}_{PR} = 1 + \frac{\mu_{SN}}{\lambda_{PR}^{F}} \tag{17}$$

### 5.2 Analysis of the Handover Latency

In this discussion we are going to explain the handover latency that would be the time required among the last moment the MU might send and receive packets through repoted server and the first time it might send and receive packet by mix zones[23,41]. So by DMM protocol as defined in [15] , the procedure of handover might be composed of the following three phases (see Figure 8):

$$T_{HL}^{DMM} = T_{L2} + T_{Auth} + T_{Binding}^{DMM} \tag{18}$$

As soon as MU will transfer from the permises of one mix zone to another mix zone , it will execute the handover procedure by starting layer 2 and authinitcation process.So this mechanism belong to wireless communication and authenticationm mechanism will be used. Therfore , $T_{L2}$, that might be the elapsed time for new Layer 2 link formation, and $T_{Auth}$, that might be the elapsed time to approve a MU, are going to be assumed constant in DDMM and FDMM.

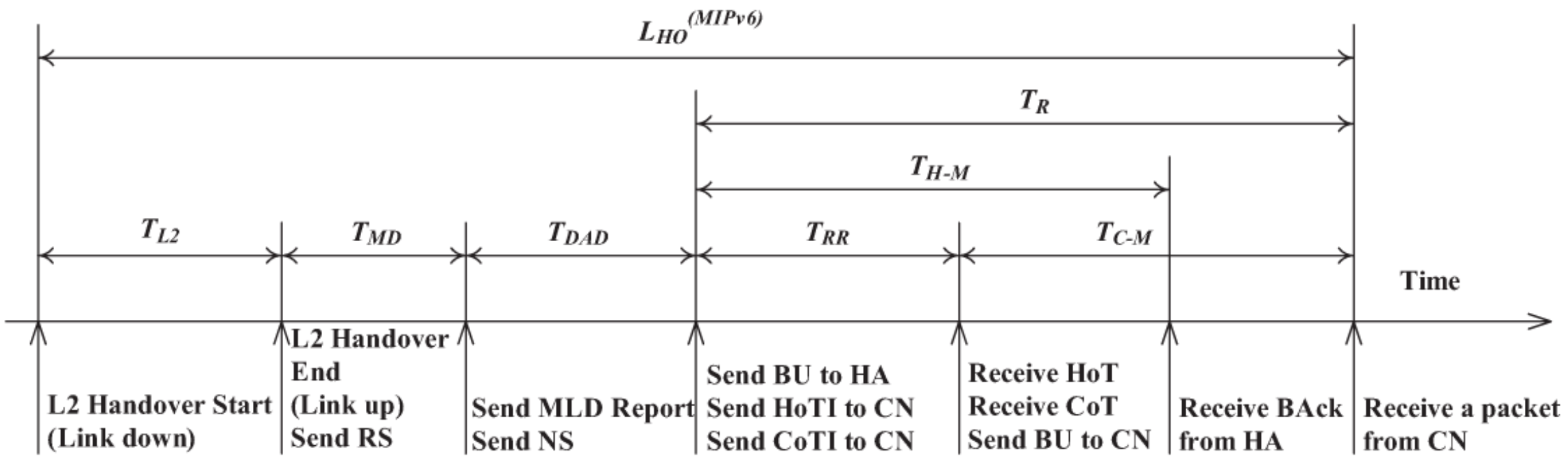

Figure 8: DDMM Handover Operation Timeline

$T_{Binding}^{DDMM}$is the intervened time of the mobility obligatory phase. The obligatory phase is necessary to apprise the MU location and recover the continuing IP flows. $T_{Binding}^{DDMM}$is given by

$$T_{Binding}^{DMM} = T_{MD}^{DDMM} + T_{LU}^{DDMM} \tag{19}$$

$T_{MD}^{DDMM}$characterizes the movement detection latency,It is articulated as follows:

$$T_{MD}^{DDMM} = 2d_{MU-MZ}(c) \tag{20}$$

where$d_{MN-MZ}(c)$ is:

$$d_{MU-MZ}(c) = \left(\frac{L_c}{BW_w} + l_w\right)\left(\frac{1}{1-P_f}\right) h_{MU-MZ} \tag{21}$$

$T_{LU}^{DDMM}$comprises ,swapping PBU/PBA messages and the handling time taken for updating the MU location and creating the tunnel.

$$T_{LU}^{DDMM} = \ 2d_{LBS-MZ}(c) + T_{PC}^{LBS} + 2T_{PC}^{MZ} \tag{22}$$

where$d_{LBS-MZ}(c)$is:

$$d_{LBS-MZ}(c) = \left(\frac{L_c}{BW} + l\right) h_{LBS-MZ} \tag{23}$$

Apart from that , FDMM might work as pre-FDMM or re-FDMM, it totally depends upon the conditions . Moreover as we are going to compare it with DDMM , it is concluded that two modes of FDMM might need extra signalling cost because of the handover procedure for the MU. It is very much obvious from figure 9 , if $t \leq x$, so the handover procedure will be executed in reactive mode, in other condition it might be performed in predictive mode.The following values are being used for FDMM performance analysis.

$x$: represents time span required when RSS will reaches to particular upper limit until and unless MU scns its nearest Access Networks.

$T$: represents time span when an L2 might sent information to the repoted server about when the MU would receive handover signal from the repoted server

$t$: represent time span about when the RSS would reaches upto maximum limit until and unless L2 link down event.

$\emptyset$: represent time span from when an L2 information would sent to repoted server to about L2 link goes down $x < \emptyset \leq \mathrm{x} + T$ $T_{HI}$: This might be the time require for the handover procedur initiation process, which might include HI and Hack message exchange.

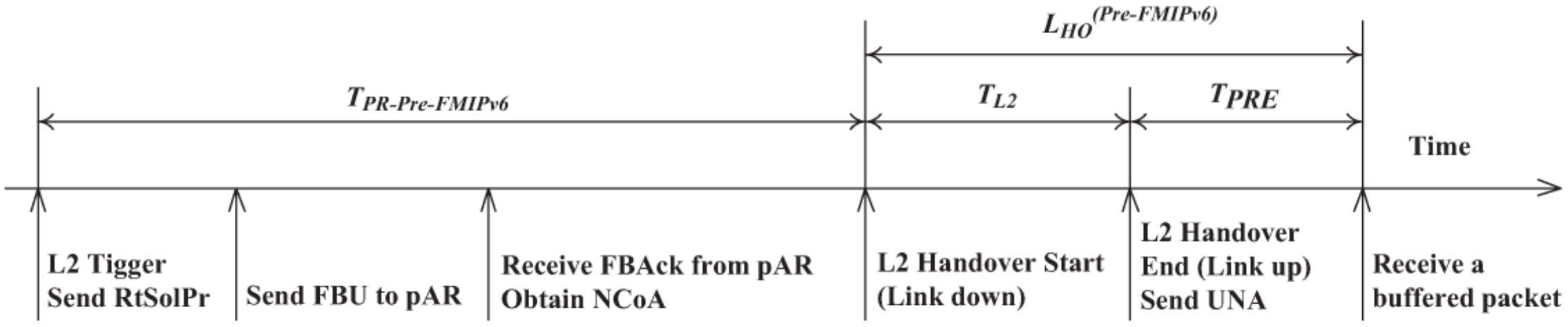

Figure 9: Pre-FDMM Handover Operation Timeline

So the pre-FDMM procedure may executed once $t > x$. So the latency for handover that is for pre-FDMM might be differ by the time at which the link L2 will goes down.So there might be the possibility of two cases : If $\emptyset = \mathrm{x} + T$, the MU's *L2* link with repoted server goes down by acknowledging the Handover command. During this procedure , the mix zones server have previously done with the pre handover procedure and created a tunnel with repoted server, however if $x < \emptyset < x + T$, the MU's *L2* link with repoted server might go down afore getting the handover command.So , now it has been concluded that the prehandover process and time required for tunnel establishment between mix zones server and repoted server might be appended in handover latency. Because $T = 2d_{MU-MZ}(c) + T_{HI}$, it can be written as the pre-FDMM handover latency as follows:

$$T_{HL}^{pre-FDMM} = \Big(\big((2d_{MU-MZ}(c) + T_{HI}) - \emptyset\big) + \big((T_{L2} + T_{Auth}) - x\big)\Big) \tag{24}$$

However x might be accomplished through the pre-handover procedure, one can not append it during the L2 link establishment and authentication phase. The $T_{HI}$is articulated as follows:

$$T_{HI} = 2d_{MZ-MZ}(c) + T_{PC}^{MZ} \tag{25}$$

where$d_{MZ-MZ}(c)$is:

$$d_{MZ-MZ}(c) = \left(\frac{L_c}{BW} + l\right) h_{MZ-MZ} \tag{26}$$

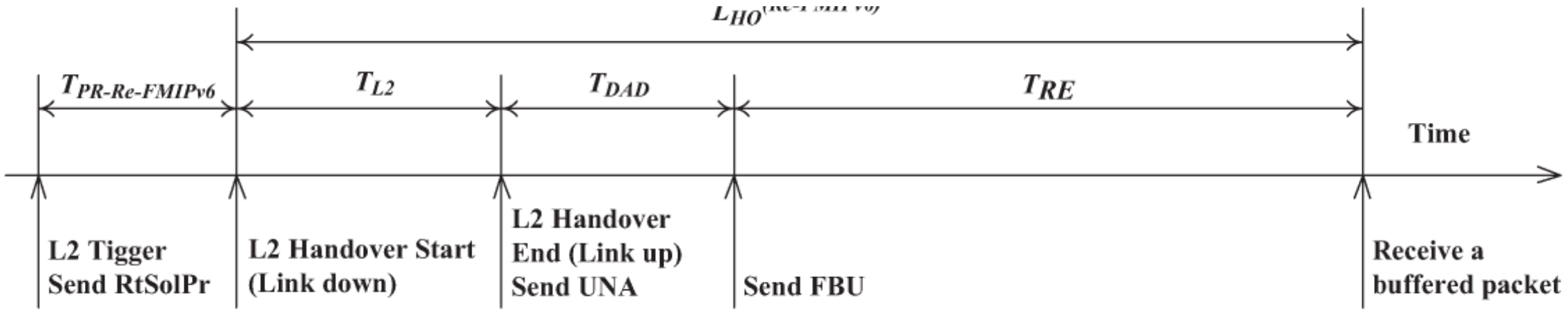

Figure 10: re-FDMM Handover Operation Timeline

It can also be written the latency for re-FDMM handover, as represented in Figure 10, $T_{HL}^{re-FDMM}$ as

$$T_{HL}^{re-FDMM} = T_{L2} + T_{Auth} + T_{HI} \tag{27}$$

## 5.3 Failure Probability Analysis of the Handover

So Handover Failure Probability might be defined as the probability that the sub-network/mix zone residence time might be less then the hand over latency [11].We can generally express the handover failure probability as follows [41]:

$$PF(.) = Prob\{T_{SN} < HL(.)\} = \int_0^{HL(.)} \mu_{SN} e^{-\mu_{SN} x}\, dx = 1 - e^{-\mu_{SN} HL(.)} \tag{28}$$

$T_{SN}$ can be explained in equation (8) by the mean subnet residence time, that might be the inverse ratio of the subnet average crossing rate. So this might be applied to DDMM and FDMM.

### 5.4 Analysis of the Session Recovery Time

So it can be defined as the time intercval from the last data packet of the session that might be acknowledged by the terminal before the procedure of handover for the first packet acknowledged after the handover [23]. So if we look into DDMM ,it is obvious that the session recovery time as depicted in figure 8 , might overlaps with handover latency and the time required for MU to acknowledge the first data packet.

$$T_{SR}^{DDMM} = \left(T_{HL}^{DDMM} - d_{MU-MZ}(c)\right) + d_{MZ-MZ}(d) + d_{MU-MZ}(d) \qquad (29)$$

Where $d_{MZ-MZ}(d)$is:

$$d_{MZ-MZ}(d) = \left(\frac{L_d}{BW} + l\right) h_{MZ-MZ}$$

So if we consider FDMM, session recovery time in FDMM might get equivalent to the handover latency and time from repoted server to mix zones server ,and from mix zones server to MU.So we can define session recovery time in re-FDMM as follows:

$$T_{SR}^{re-FDMM} = T_{L2} + T_{Auth} + T_{HI} + d_{MZ-MZ}(d) + d_{MU-MZ}(d) \qquad (30)$$

Apart from that ,

$$T_{SR}^{pre-FDMM} = \left(max\{(2d_{MU-MZ}(c) + T_{HI}) - \emptyset, 0\} + \left((T_{L2} + T_{Auth}) - x\right) + d_{MU-MZ}(d)\right) \qquad (31)$$

As revealed in Figure 9,$\left(\mathrm{d}_{\mathrm{MU-MZ}}(c) + \left((\mathrm{T}_{\mathrm{L2}} + \mathrm{T}_{\mathrm{Auth}}) - \mathrm{x}\right)\right) > \mathrm{d}_{\mathrm{MZ-MZ}}(\mathrm{d})$, we might incorporate the time when $L2$ link swapped among ANs during the session recovery time.Moreover, the reported server would initiate sending the data traffic towards mix zones server as soon as it obtains the Hack message and prior directing handover command. During this ,It is depicted that the MU will send and receive data through reported server until and unless it receive the handover command.Therfore the time required to direct handover command might not incorporated in the session recovery time.

### 5.5 Analysis of the Packet Loss

It can be defined as the total number of data traffic loss or we can say packet drop in between the MU handover.As it can be seen that there might not be any buffering mechanism will be used during the handover procedure [23].So we might assume that packet transfer rate to and from MU will be $\lambda = \lambda_p \bar{N}_{PR}$ packets/second, where $\lambda_p$can be the packet rate per active prefix. So we can express packet loss as follows:

$$PL_{DMM} = \lambda L_d T_{SR}^{DDMM} \qquad (32)$$

In the same way , the packet loss in re-FDMM is characterized, as:

$$PL_{re-FDMM} = \lambda T_{SR}^{re-FDMM} \qquad (33)$$

Apart from that,the mechanism of buffering is being active in pre-FDMM, therefore,it can be concluded that the information loss will be directly proportional to the tunnel forming interval , As the MU is being disconnected from the reported server and to the buffer overflow at the mix zones server(see Figure 9) [42, 43]. Therefore,it can be defined as the packet loss will be defined as the total of information transported afore beginning the buffering and the packets that surpass the buffering size later initiating the buffering, as follows:

$$T_{Buffering}^{FDMM} = \left(d_{MU-MAAR}(c) + \left((T_{L2} + T_{Auth}) - x\right)\right) - \left(d_{MZ-MZ}(d)\right) \tag{34}$$

$$PL_{pre-FDMM} = \lambda L_d\left(max\{(d_{MU-MZ}(c) + T_{HI}) - \emptyset, 0\} + max\{T_{Buffering}^{FDMM} - B, 0\}\right) \tag{35}$$

where $B$ is the buffer size.

## 5.6 Analysis of the Signaling Cost

So it has been deduced that signaling cost would be the combination of handover related signaling message for an MU.So , that might be computed using the product of mobility signaling message and the hop distance [44, 45, 46].However in DDMM , the cost of signaling might contain RS message among the MU and the mix zone,therefore registration of PBU/PBA handshakes among the mix zones server and the LBS server for registration of nLNP, So that the PBU/PBA message would get exchanged among every RSU and the LBS server.Therfore , Signalling cost of DDMM might be defined as:

$$C_{sig}^{DDMM} = 2\mu_{sn}L_c\left(h_{MU-MZ} + h_{LBS-MZ}(\bar{N}_{PR} + 1)\right) \tag{36}$$

As it has already been told that , vehicles moves away from mix zone, therefore it is been obvious that the number of hop count among mix zone server and RSU would get increase every time the handover executed.However by enhancing the number of hop count might increase the cost of signaling of FDMM in an exponential way.This might be known as the worst case for FDMM and has no influence on signaling cost of DDMM. Moreover signaling cost for FDMM predictive mode might consist of , DDMM signaling and the extra signaling message swapped among mix zone server.So it can be represented as follows:

$$C_{sig}^{pre-FDMM} = 2\mu_{sn}L_c\left(h_{LBS-MZ} + h_{MU-MZ} + \sum_{n=1}^{\bar{N}_{PR}} nh_{MZ-MZ}\right) \tag{37}$$

So if we consider reactive mode,there might not be any RS and RA message would get exchanged among MU and mix zone server,because the procedure of handover might starts after connectivity to the mix zone server.So the cost of signaling for re-FDMM might be :

$$C_{sig}^{re-FDMM} = 2\mu_{sn}L_c\left(h_{MZ-MZ} + \sum_{n=1}^{\bar{N}_{PR}} nh_{MZ-MZ}\right) \tag{38}$$

# 6 Numerical Results and Discussion

Now we shall discuss the results in this section .We have presented the default parameter values in table Table II, that might be obtained from Hossain, M. S. et al. [39, 40], Seonggeun, R. et al. [47], Ali-Ahmad et al. [41], and Giust, F. et al. [23].So it is clear from the table that assigned value for the layer2 latency L2 might includes scanning phase is 330 ms , however scanning phase x is 300 ms. Apart from that , $L_{Auth}$ is known to be authentication latency and has value of 100 ms. We have analysed the impact of multiple parameters over handover performance by putting different values.

Table II: Default Parameter Values

| Parameter | Value |
|---|---|
| *X, Y* | 36 &24 km |
| $S_x = S_y$ | 200 m |
| $K_1 = K_2$ | 5 |
| $U_{max}$ | 25 Sec |
| $1/\lambda^F_{PR}$ | 240 Sec |
| $L_c, L_d$ | 80 & 400 bytes |
| $BW, BW_w$ | 100 & 10 Mbps |
| $l, l_w$ | 0.5 & 2 ms |
| $T^{LBS}_{PC}, T^{MZ}_{PC}$ | 20, 10 ms |
| Ø | 35 ms |
| $B$ | 500 KB |
| $\Lambda$ | 50 packets/s |
| $\xi$ | 0.5 |
| $p_f$ | 0.5 |
| $\bar{v}$ | 25 m/s |
| $R$ | 1000 m |
| *Simulator* | SUMO[61] |

In the start , we have analyzed the effect of mix zone radius r. We have assumed all the default values excluding those values for which we have vareyd the mix zone radius from 1000 to 6000 m, by considering dissimilar transmission technique range (e.g., 802.11p, WiMax, LTE). This might fluctuate the number of mix zones from 285 to 4 using Sumo simulator.

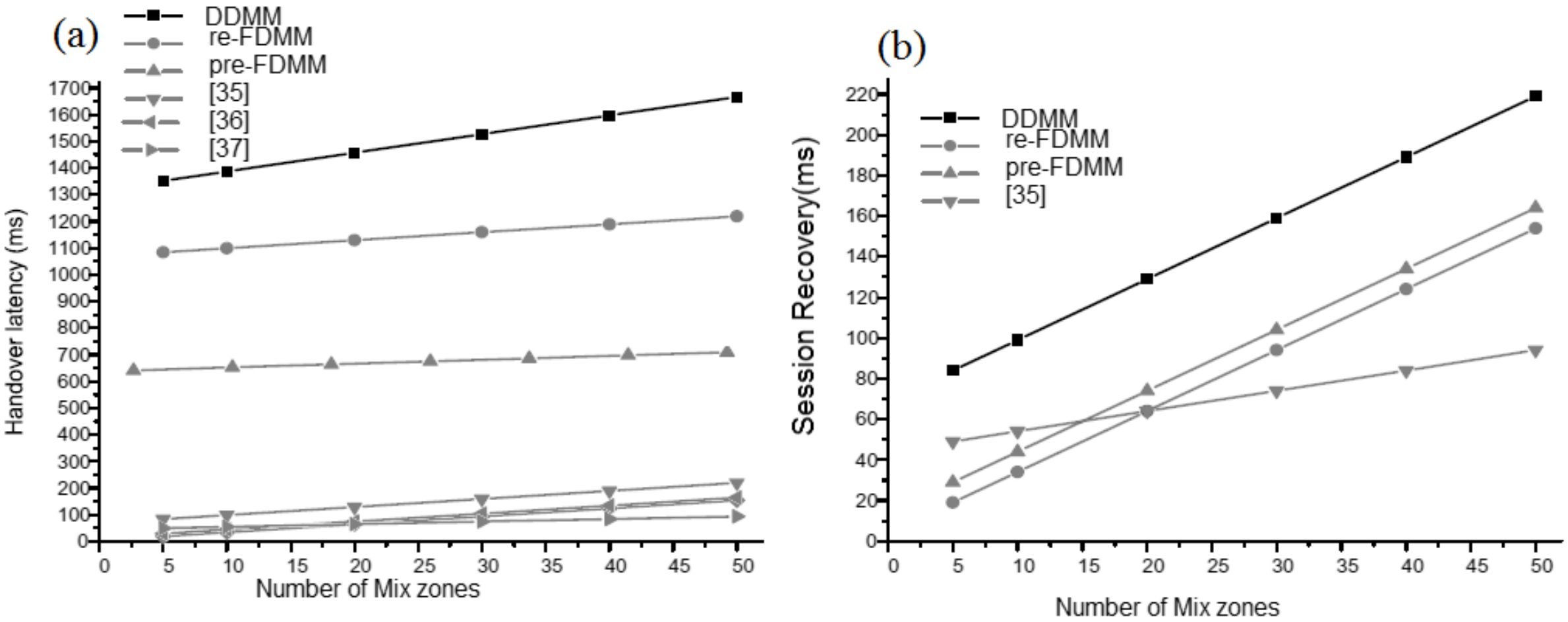

Figure 11: The impact of *r* (radius of mix zone) on handover latency and session recovery.

As we are going to enlarge the mix zone radius , consequently the number of hops among network entities might also get decreased and hence the handover latency might get affected.As represented in Figure 11(a) , the value of handover latency for DDMM and re-FDMM might get reduced a little by enlarging the mix zones radius , as the number of mix zones and LBS server , and among mix zones will get reduced.However if we consider pre-FDMM , then its obvious that there might not be signaling exchanged among network objects once the L2 link went down by acquiring handover command.So forth the process of binding resistration may occur in prior as the link went down.Therfore by changing the value of r has no any implication on the pre-FDMM handover latency and reduce handover latency [35,36,37].Moreover the session recovery time , as depicted in figure 11(b) has shown the same drift as handover latency with a little delay enhancement because of the delay in sending the data traffic to the vehicle and at the same time our DDMM method has fast session recovery as compare to re-FDMM ,pre-FDMM and also with [35- 37].

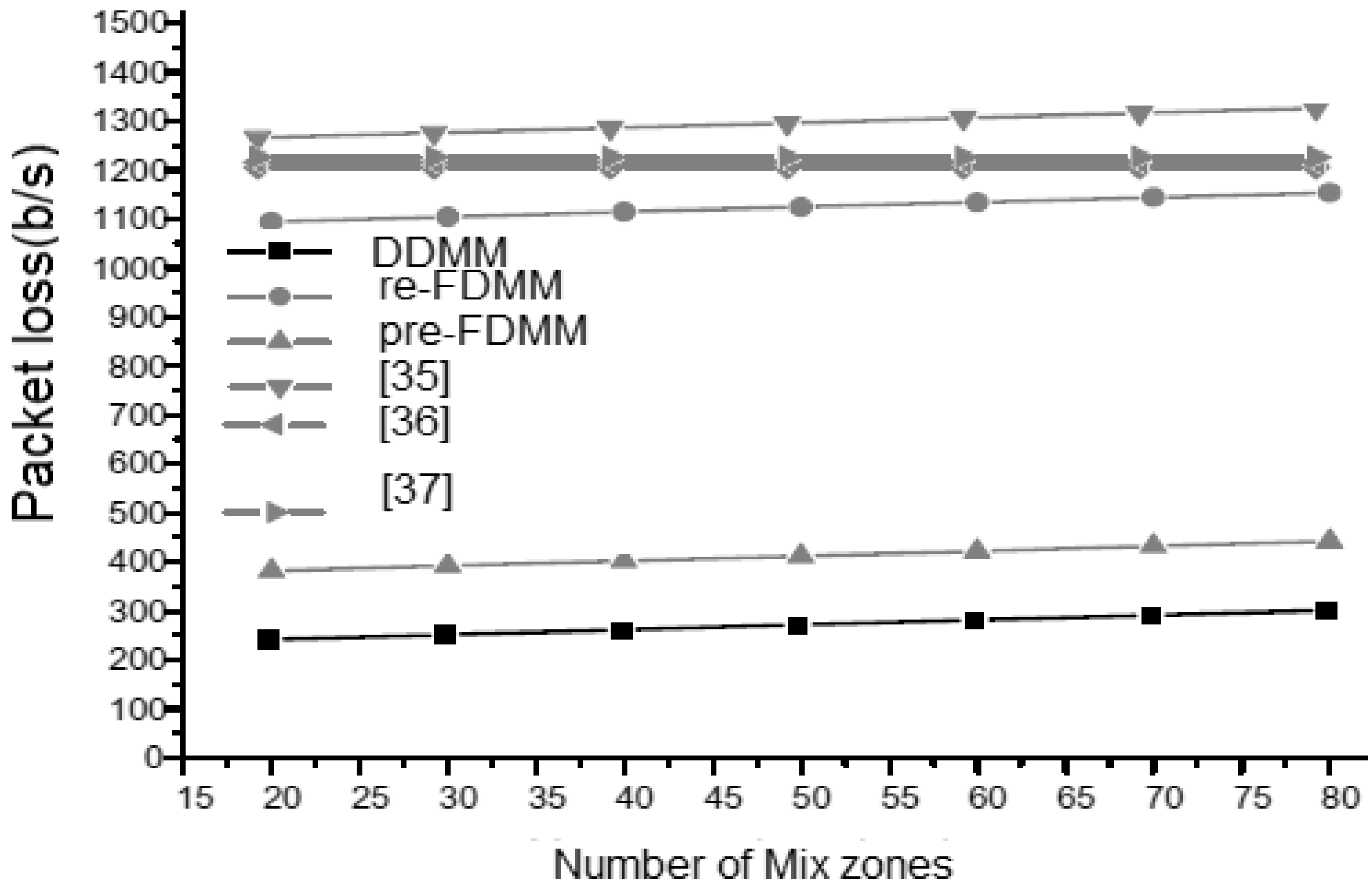

Figure 12: The impact of *r* (radius of mix zones) on packet loss.

Figure 12 is goig to show the effect of the mix zone radius on packet loss.It can be seen that the packet loss during DDMM and re-FDMM process is directly proportional to the session recovery time because there might not be any buffering procedure is used.However packet loss during DDMM and re-FDMM might get decreased as soon as the session recovery time get shorten.Moreover , during the pre-FDMM procedure , the buffering will be performed for ongoing data traffic at mix zones server and therefore loss of packet might get removed as long as the data traffic might not get surpass the buffering size.

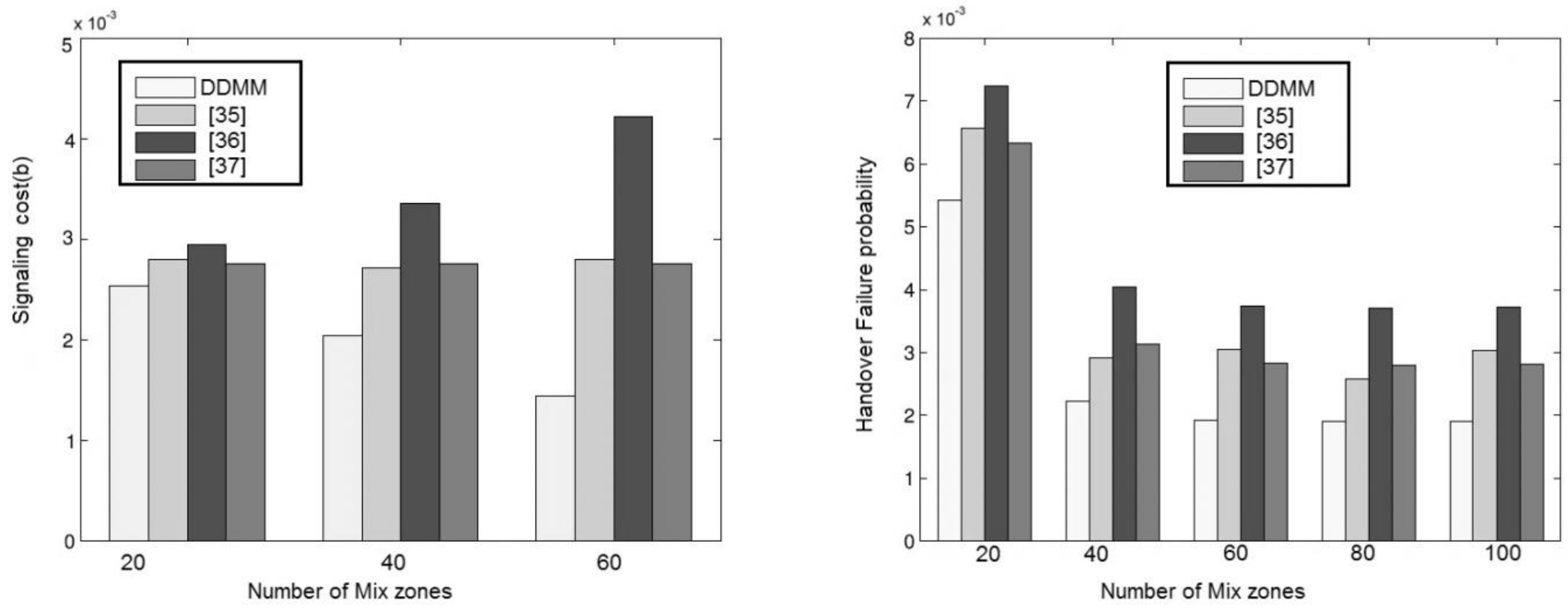

Figure 13: The impact of *r* (radius of mix zones) on handover failure probability and signaling cost.

By considering figure 13 , it can be analysed that by enlarging the mix zone radius and by considering constant vehicle speed might reduce the subnet crossing rate $\mu_{SN}$. Therfore , the probability of handover failure might get decreased because it might have direct proportion to subnet crossing rate. Moreover , due to low crossing rate the number of handover events might also decreased ,therefore,the cost of signaling might get decreased in DDMM as compred to previous method[35,36,37].

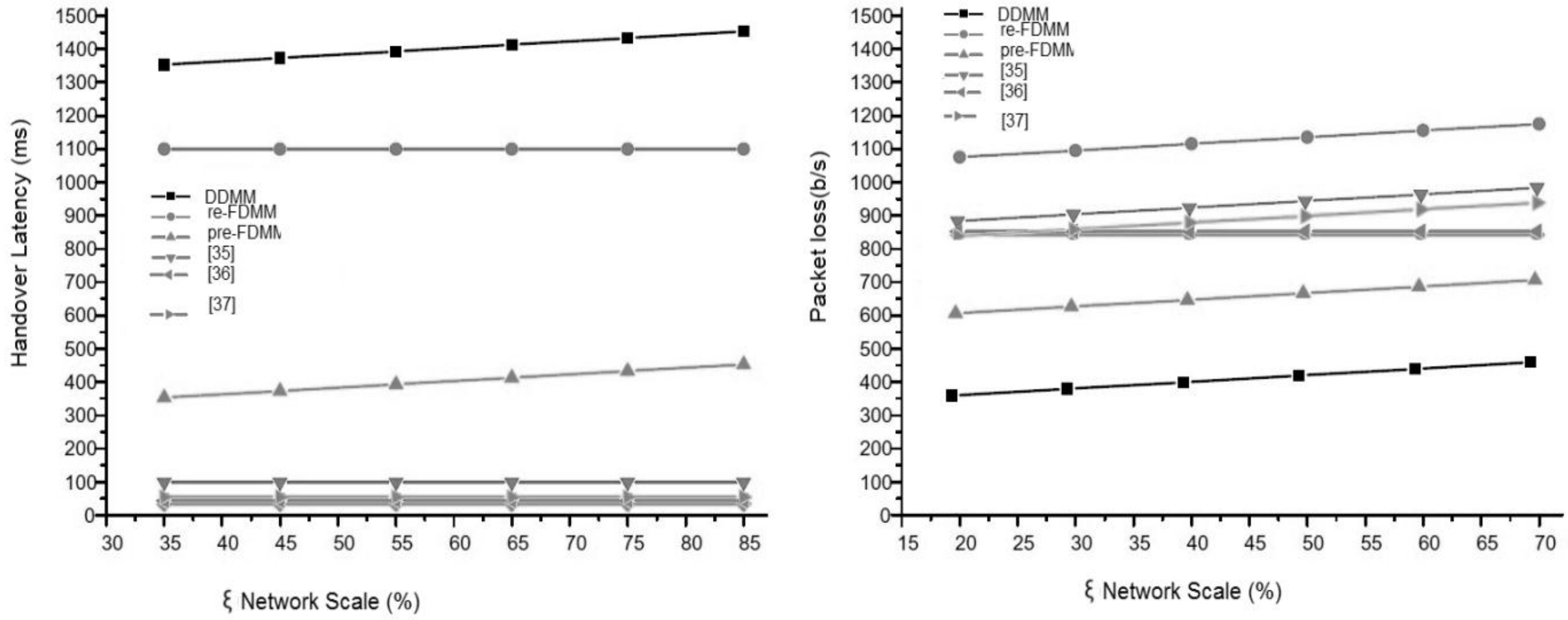

(a) (b)

Figure 14: The impact of $\xi$ (network scale) on handover latency and packet loss.

It is also observed that wireless and wired signaling might also effect the handover performance. We have interogate about the wired link effects by analyzing the network scaling impact . We have shown variation in network scale $\xi$ from 0.1 to 1,that means we have varied the ratio of the distance among mix zones server to the distance among mix zone and LBS server.

It is depicted in figure 14(a) regarding the effect of network scale over handover latency and packet loss.So it has been analysed that there might not be any effect on the handover latency of DDMM and pre-FDMM.Therefore the handover management message in DDMM might get exchanged among mix zone server and LBS server , as the number of mix zones remain constant during the network scale change process.Moreover , during pre-FDMM exchange process , the handover management message would exchanged before the L2 links goes down.Apart from that , during pre-FDMM handover procedure latency might get enhance linearly because of increment of hops counts in between mix zones during exchanging HO-initiate process message.However it is also been analysed that the network scale has the same effect on hand over failure probability because of the direct proportion among handover latency and handover failure probability.

As depicted in figure 14(b) , packet loss during DDMM and pre-FDMM enhance gradually as a result of increasing network scale ,however during pre-FDMM process buffering will remove the packet loss like previoys method [35- 37].It has been analysed that DDMM process has the greatest packet loss because of high handover latency and the transmission delay operated by reported server to mix zone, and also from mix zones server to the vehicle.Therfore by increasing the network scale might enhance the number of hop counts among reported server and mix zones server.It can also be observed that the data transmission delay among mix zone has a very important effect in pre-FDMM, the reason is that , there might be more messages passes through the hop among mix zones server to retrieve active session.So it has been deduced that by increasing network scale would enlarge the packet loss in pre-FDMM.

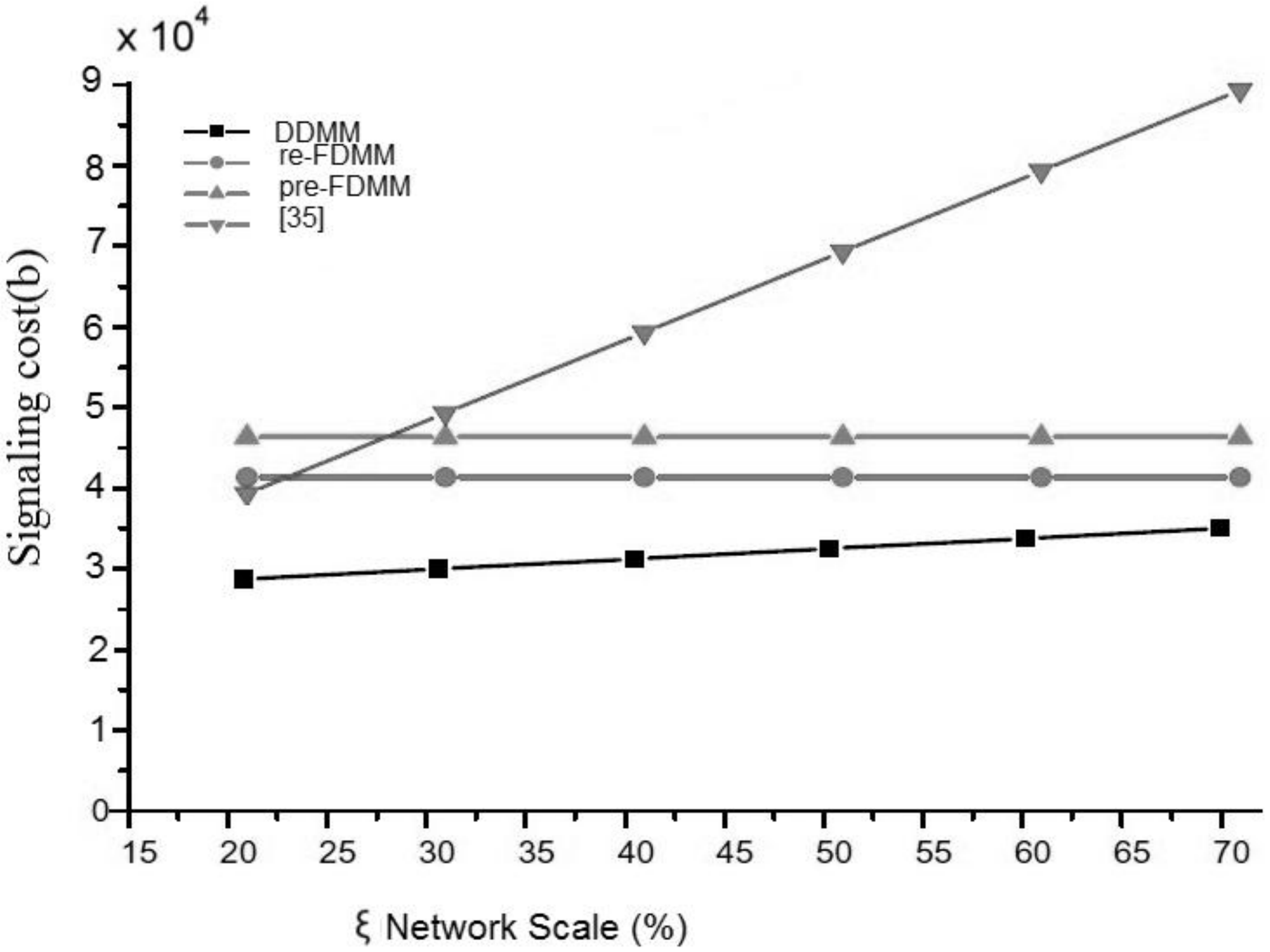

Figure 15: The impact of $\xi$ (network scale) on signaling cost.

Therfore it is also been observed that the cost of signaling for our given solution might get increase slowly because of the network scale enhancement.It is because , there might be signaling exchanged among mix zone during the handover procedure in our given solution , however this would not be the case if we consider DDMM. So it has been deduced that , the placement of our given solution is more reliable in large operator network , in which the distance among MARRs and LBS server have to be large as compared with the distance among mix zones henceforth the cost of signaling on the network performance have no any significance shown in figure. 15

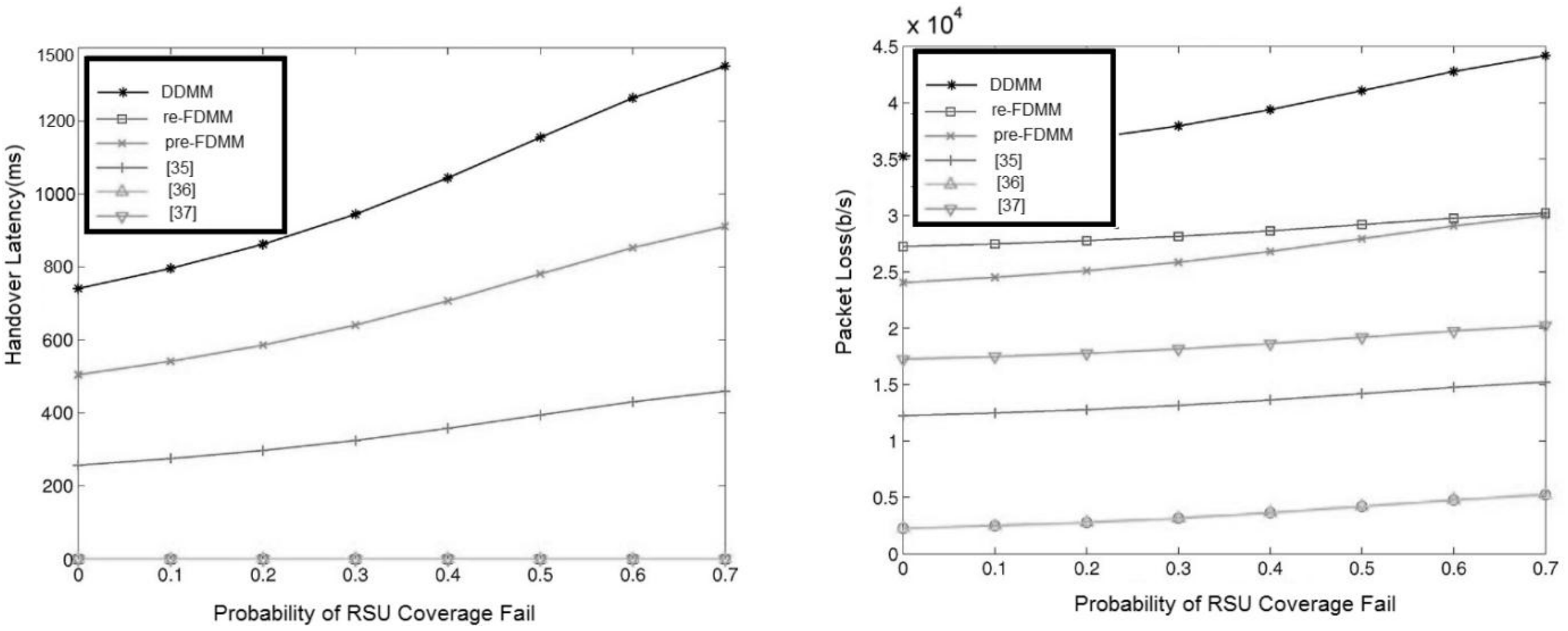

(a) (b)

Figure 16: The impact of probability of $p_f$(RSU coverage fail) on handover latency and packet loss.

Priviously we have analysed the impact of signaling over wired link during handover performance , while we also have studied the impact of signaling over RSU coverage also.Therfore we have analysed the effect of wireless link failure probability $p_f$ on handover performance.The$p_f$value might get varied from 0.1 to 0.8.

So during FDMM procedure , handover might get start after vehicle get disconnected in pre-FDMM or after recieveing a handover command from mix zones server in pre-FDMM.So that’s why it has been observed that there is no message transferred over a wireless link among vehicle and mix zone during the handover procedure , therefore the latency for handover remain constant as depicted in figure 16(a).Apart from that if we consider DDMM procedure , then the handover latency would decrease a little because of HS and HA message transfer among mix zones and the MU.

As discussed earlier , packet loss might depends upon the session recovery time until and unless buffering procedure is not installed.However there is a slight increase in session recovery time of DDMM and FDMM as the probability of RSU coverage failure might get enhance because of the RSU coverage link impact during the procedure of sending data traffic to the vehicle.So it has been analysed and depicted in figure 16(b) ,the packet loss of DDMM and re-FDMM enhanced marginally because of session recovery time increment.

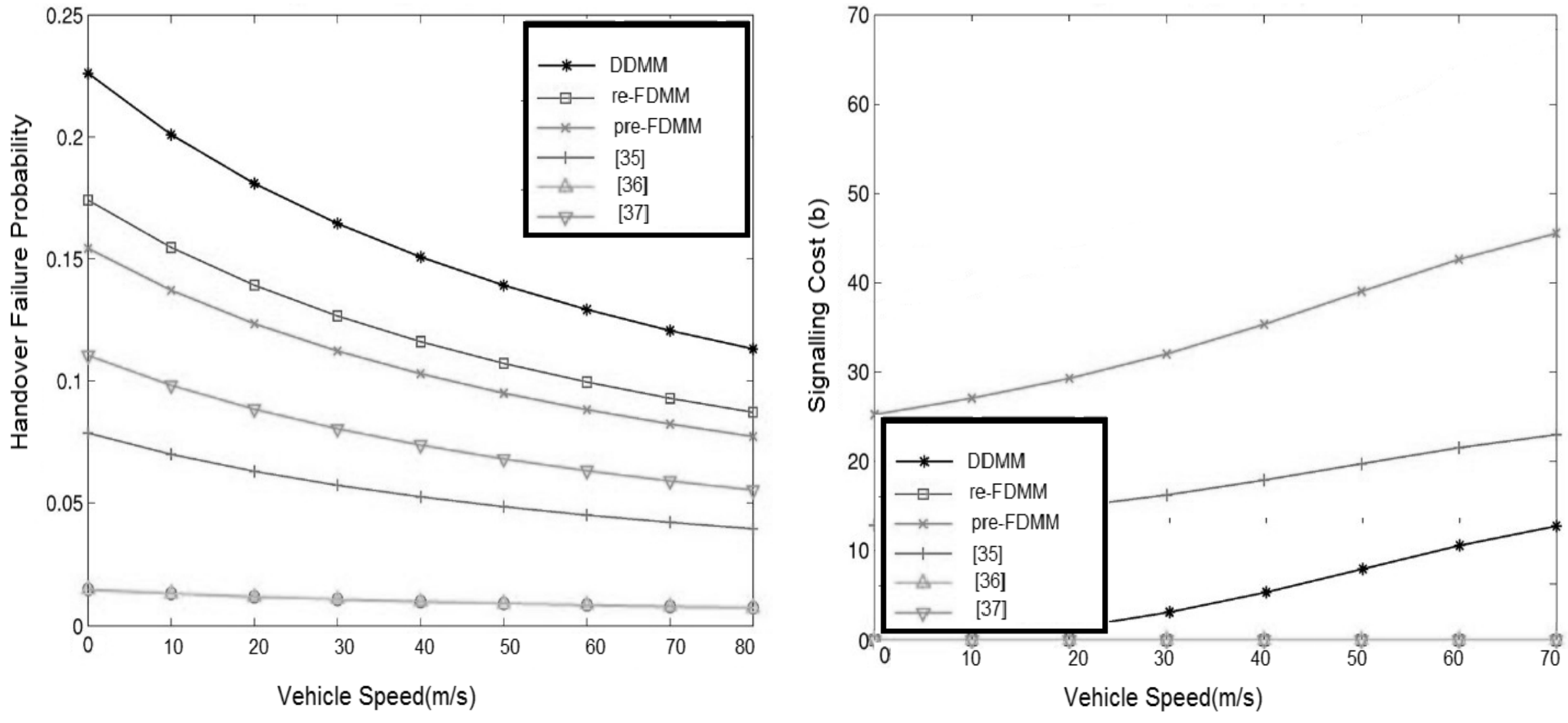

(a) (b)

Figure 17: The impact of $\bar{v}$ (vehicle speed) on handover failure probability and signaling cost.

If we consider the road network scenario , the speed of vehicle might vary suddenly based upon the environmental limitations.We have analysed the effect of speed of vehicle over handover performance by changing speed value from 0 to 100 m/s, which might be equivalent to 0 to 360 KM/H.So , this might exhibit low , medium and high speed situation (i.e., pedestrian speed, vehicle speed in urban and highway environments, and high-speed train).

We analysed the trend of handover probability and cost of signaling,with respect to vehicles speed impact, which is analogous to the mix zone radius , because of vehicles high speed might increase subnet crossing rate $\mu_{SN}$ and moreover high handover failure probability. As it is depicted in figure 17(a), the probability of handover failure for DDMM ,FDMM,and [35-37] may uplift slowly as the speed of vehicle might increases.So it can be seen from figure that crossing rate is been highest in the right most graph (a),therefore it can be deduced that probability of handover failure might have a direct relation with the handover latency however if the handover latency is small then the chances for handover failure is also low.Therfore probability for handover failure for re-FDMM is little lower then DDMM and if we are considering pre-FDMM then it has the lowest handover probability failure. So as depicted in fig 17 (b) , we have analyzed the influence of vehicle speed on the signaling cost.So it is been deduced that more handover will be performed if the vehicle speed is bit high.Moreover the handover management message are being transferred among network objects per handover event therefore signaling cost will also be increased exponentially as the vehicle speed is going to increase.So , it is been analysed that our FDMM technique will costs more signaling then DDMM technique, therefore it is been deduced that vehicle’s high speed would effect the signaling cost of our proposed solution.

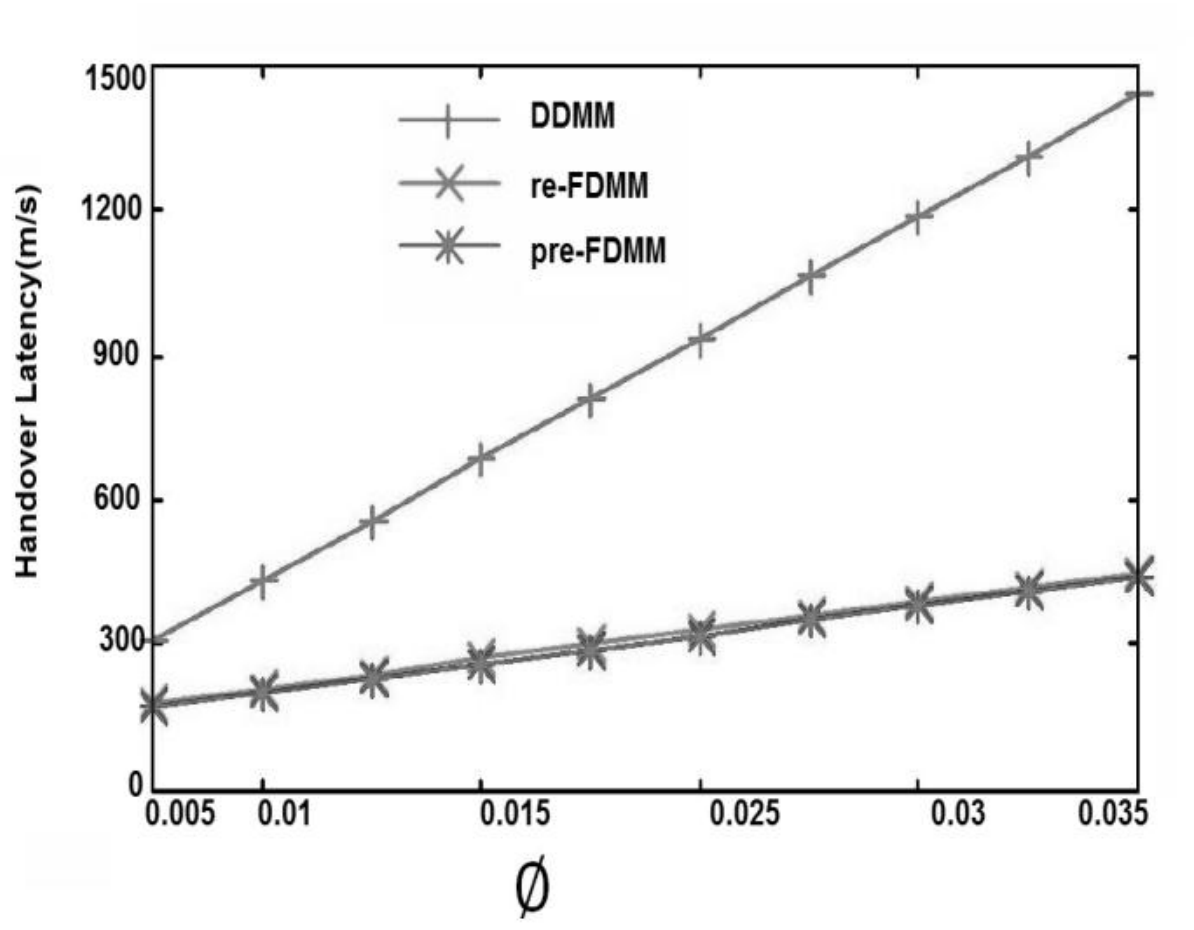

(a)

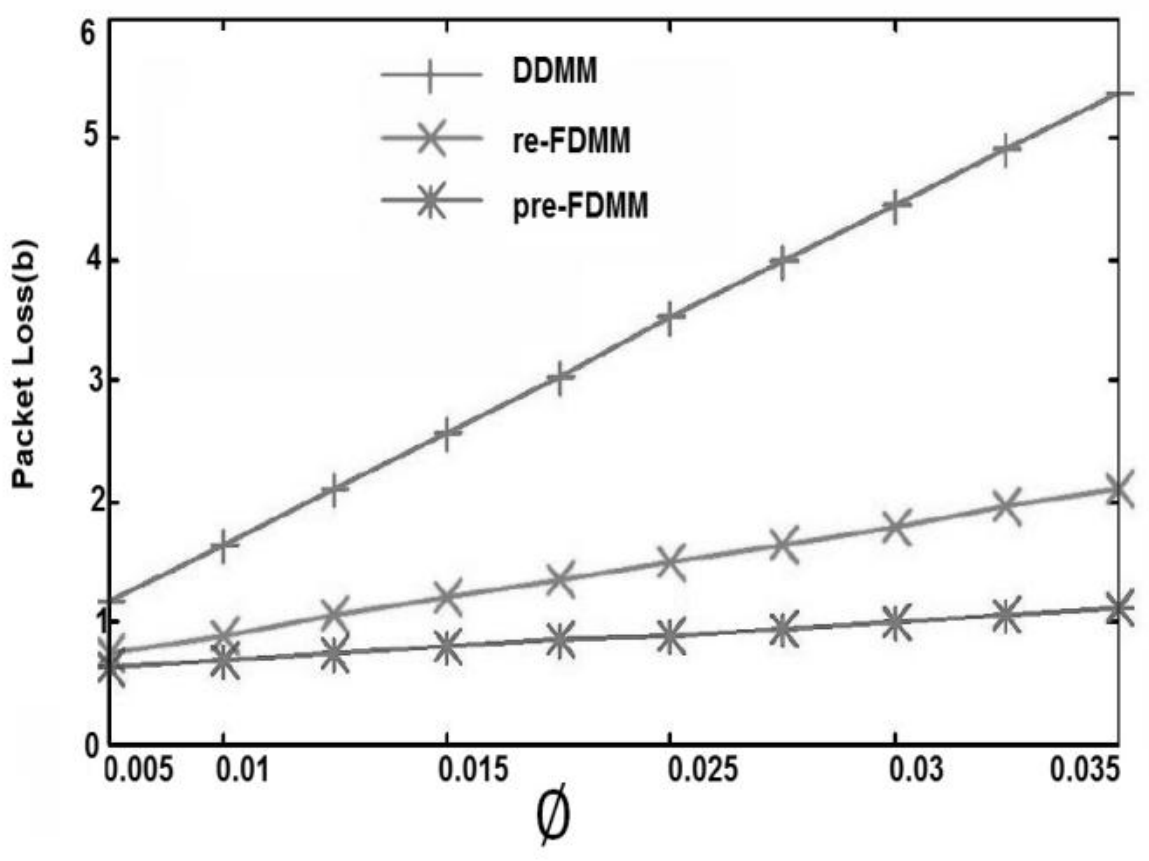

(b)

Figure 18: The impact of ∅(time between receiving L2 report till link goes down) on handover latency and packet loss.

So to judge the effectiveness of the proposed FDMM predictive mode , we analyzed the probability that the L2 link might go down before accepting the handover command.Therfore , reported server might have information regarding mix zones server by using report message and might initiate the HO-initiate procedure.So , FDMM predictive mode will utilize around 0.035 seconds from accepting L2 report until and unless directing the handover command to the vehicle. (i.e., performing the HO-Initiate process).Therfore we might utilize the parent value except for ∅ , in which we might fluctuate its value from 0.005 to 0.035 seconds.As depicted in fig 18(a) the handover latency for pre-FDMM would get increase a little as the L2 links went down before accepting the handover command (i.e., increasing the ∅ value),although it has no any implication on DDMM and re-FDMM because there might not be any exchanged message before L2 goes down.However probility for handover failure might get effected as on handover latency because of direct proportion among them.As depicted in figure 18(b) , as soon as the L2 link goes down before accepting the handover command , moreover in pre-FDMM process there is a chance of occurring packet loss because the buffering mechanism has not been commenced yet.

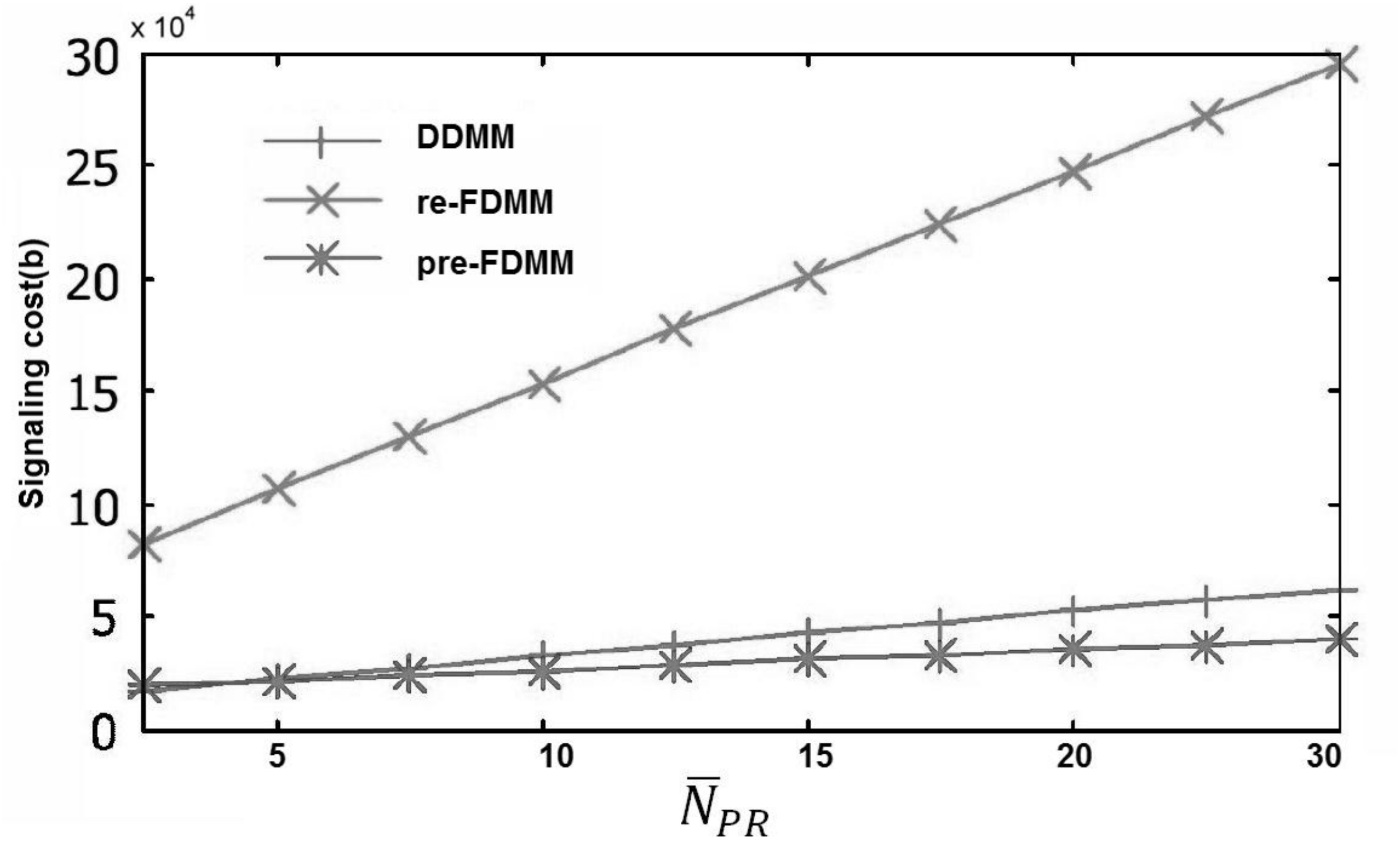

Figure 19: The impact of $\bar{N}_{PR}$on signaling cost

We have analyzed the effect of number of active prefixes $\bar{N}_{PR}$ by changing the prefix number from 4 to 29 prefixes.We assumed all the default values excluding the value of the mean of the prefix lifetime in a foreign network $1/\lambda_{PR}^{F}$.It has been deduced that by increasing the value of $1/\lambda_{PR}^{F}$. Would get increase the number of active prefixes collectively per handover event. The value of $1/\lambda_{PR}^{F}$ get fluctuate from 225 to 1500 seconds as depicted in figure 19,so it is clear that DDMM and FDMM are flow based and therefore the cost of signaling might get increase as the active prefixes get increase.As it was discussed earlier , as the vehicle is getting away from mix zone . Henceforth , as the number of active prefixes getting high as the number of hops among mix zone server and RSU increases per every handover.Apart from that , total number of hops among all mix zones and LBS server remain contant as it is already been assumed that the variances are very small.According to DDMM , it is been deduced that number of mix zonesexchange the handover management message with LBS server by rising the number of active prefixes. Therfore the cost of signaling might rise in a linear way ,while the number of hop counts among all mix zones and LBS server remain constant.According to FDMM , signaling cost for both modes has to be changed similalrly with a bit higher cost in pre-FDMM because of message transfer among MU and mix zone.Nothing like DDMM , it is observed that handover management messages are being transferred among mix zones rather then LBS server.Therfore , the cost of signaling of FDMM is bit greator then the DDMM by exponential increment drift to the because of the rise of swapped number of handovers management messages and the number hops among mix zones per handover.

## 7 Conclusion

We have proposed dynamic distributed mobility management( DDMM) method based on PFMIPv6.We have introduced two operation modes as defined in PFMIPv6: reactive mode and predictive mode. FDMM shortens the handover latency by performing movement detection and binding registration process , moreover we compared our schemes with previous schemes [35,36,37]. In further, we have proposed a buffering mechanism to reduce packet loss by buffering packets intended for the vehicle at mix zones server.We have conducted an analytical evaluation of FDMM and other methods and compared results with DDMM. We have also derived analytic formulas for the handover latency, handover failure probability, session recovery time, signaling cost and packet loss.Simulation results have revealed that DDMM scheme has outperformed FDMM with an extra signaling cost.